\documentclass[aps,pre,bibnotes, superscriptaddress]{revtex4}

\usepackage{graphicx}
\usepackage{amssymb}
\usepackage{amsmath}
\usepackage{amsfonts}
\usepackage{color}
\usepackage{hyperref}
\usepackage{bm}

\allowdisplaybreaks

\DeclareMathAccent{\ring}{\mathalpha}{operators}{"17}
\providecommand{\st}[1]{_{\text{#1}}}

\providecommand{\ut}[1]{^{\text{#1}}}
\providecommand{\ten}[1]{{\bv{#1}}}

\def\onehalf{\frac{1}{2}}

\def\bra{\ensuremath{\langle}}
\def\ket{\ensuremath{\rangle}}

\def\pd{\partial}

\def\Imat{\mathbb{I}}
\def\im{\mathrm{i}}

\def\drho{\delta \rho}

\def\kv{\bv{k}}
\def\qv{\bv{q}}
\def\uv{\bv{u}}
\def\jv{\bv{j}}

\def\vv{\bv{v}}

\def\rv{\bv{r}}
\def\hv{\bv{h}}
\def\b0{\bv{0}}

\def\kt{\tilde{k}}
\def\qt{\tilde{q}}

\def\drho{\delta\rho}
\def\RE{\text{Re}}
\def\IM{\text{Im}}

\def\ra{\rightarrow}

\def\Fcal{\mathcal{F}}

\newcommand{\bitem}{\begin{itemize}}
\newcommand{\eitem}{\end{itemize}}
\newcommand{\benum}{\begin{enumerate}}
\newcommand{\eenum}{\end{enumerate}}
\newcommand{\bblock}[1]{\begin{block}{#1}}
\newcommand{\eblock}{\end{block}}
\newcommand{\bmini}[1]{\begin{minipage}{#1}}
\newcommand{\emini}{\end{minipage}}
\newcommand{\btab}[1]{\begin{tabular}{#1}}
\newcommand{\etab}{\end{tabular}}
\newcommand{\btabn}[1]{\begin{tabular}{#1}}
\newcommand{\etabn}{\end{tabular}}
\newcommand{\beq}{\begin{equation}}
\newcommand{\eeq}{\end{equation}}
\newcommand{\beqn}{\begin{equation*}}
\newcommand{\eeqn}{\end{equation*}}
\newcommand{\bmult}{\begin{multline}}
\newcommand{\emult}{\end{multline}}
\newcommand{\bsplit}{\begin{split}}
\newcommand{\esplit}{\end{split}}

\newcommand{\bv}[1]{\mathbf{#1}}

\begin{document}
 \title{Critical dynamics of an isothermal compressible non-ideal fluid}
 \author{Markus Gross}
 \email{markus.gross@rub.de}
 \affiliation{Interdisciplinary Centre for Advanced Materials Simulation (ICAMS), Ruhr-Universit\"at Bochum, Universit\"atsstr. 90a, 44789 Bochum, Germany}
 \author{Fathollah Varnik}
 \affiliation{Interdisciplinary Centre for Advanced Materials Simulation (ICAMS), Ruhr-Universit\"at Bochum, Universit\"atsstr. 90a, 44789 Bochum, Germany}
 \affiliation{Max-Planck Institut f\"ur Eisenforschung, Max-Planck Str.~1, 40237 D\"usseldorf, Germany}

\begin{abstract}
A pure fluid at its critical point shows a dramatic slow-down in its dynamics, due to a divergence of the order-parameter susceptibility and the coefficient of heat transport. Under isothermal conditions, however, sound waves provide the only possible relaxation mechanism for order-parameter fluctuations. 
Here we study the critical dynamics of an isothermal, compressible non-ideal fluid via scaling arguments and computer simulations of the corresponding fluctuating hydrodynamics equations. 
We show that, below a critical dimension of 4, the order-parameter dynamics of an isothermal fluid effectively reduces to ``model A,'' characterized by overdamped sound waves and a divergent bulk viscosity. In contrast, the shear viscosity remains finite above two dimensions. 
Possible applications of the model are discussed.
\end{abstract}


\maketitle

\section{Introduction}
The presence of long-range static correlations can induce drastic slowing down of the dynamics of a fluid at the critical point \cite{stanley_book, hohenberg_halperin}.
This can be understood by noting that the order-parameter relaxation rate is typically given as a ratio between a transport coefficient and a susceptibility: the susceptibility diverges critical point, while the transport coefficient remains roughly constant (or has a much weaker divergence).
The characteristic dependence of the relaxation rate $\Gamma$ on the wavenumber, $\Gamma\propto k^z$, or, by virtue of the dynamic scaling assumption, equivalently on the correlation length, $\Gamma\propto \xi^{-z}$, defines the \textit{dynamic critical exponent} of the order parameter, $z$.
In addition to the order-parameter relaxation time, other transport coefficients of a fluid, such as the viscosity, are often divergent as well and entail their own critical exponents.
Similarly to statics, many properties of dynamic critical fluctuations, such as dynamic critical exponents or amplitude ratios, are universal and thus not specific to a particular substance.
For instance, an energy-conserving pure (i.e., single-component) fluid at the liquid-vapor critical point and a binary fluid at the demixing point both belong to the same dynamic universality class of \textit{model H} and thus share the same set of dynamic critical exponents \cite{hohenberg_halperin}.

While the conventional pure and binary fluid have been extensively investigated both by theory and experiments (see, e.g., \cite{hohenberg_halperin, sengers_supercritical_1994, folk_moser_review2006, taeuber_book, onuki_book} for reviews), the critical dynamics of an \textit{isothermal} compressible, single-component fluid seems not to have received much attention so far.
Two-dimensional isothermal fluids are often employed, for instance, as simple models for monolayer films that are confined to liquid interfaces \cite{schwartz_flow_prl1994, stone_film_physfl1995, lubensky_goldstein_physfl1996}. Indeed, many of these films are also known to undergo liquid-vapor-like phase transitions \cite{gaines_monolayers_book_1966, aratono_phase_films_jcoll1984, middleton_pethica_monolayers_prsa1984, rasing_kim_ellipsometry_pra1988, knobler_sci1990, kaganer_monolayer_review1999, casson_lg_monolayer_1999, tomassone_koplik_surfactants_langm2001, varga_liquidgas_jphyschem2005}.
Recently, there has been growing interest in understanding the critical properties of these and related lipid bilayer systems \cite{keller_mcconnell_prl1999, nielsen_nat2000, nielsen_langm2007, veatch_keller_prl2002, murtola_bilayers_prl2006,veatch_mixtures_pnas2007, honerkamp_smith_bioj2008, ehrig_schwille_membranes_NJP2011, honerkamp_smith_prl2012, machta_casimir_cells_prl2012}, especially, since they constitute the building blocks that form the membranes of biological cells \cite{mouritsen_lipidomics_book}. 
Isothermal non-ideal fluid models have also been used to study phase-separation \cite{langer_turski_pra1973, valls_mazenko_prb1988, farrel_valls_prb1989, farrel_valls_prb1990, farrel_valls_prb1991, osborn_spinodal_1995, lamorgese_mauri_physl2009}, capillary waves \cite{felderhoff_physica1970, turski_langer_pra1980,shang_jcp2011} and supercooled liquids close to the glass transition \cite{das_mazenko_prl1985, das_mazenko_pra1986, kirkpatrick_mct_pra1986a, kirkpatrick_mct_pra1986b, das_book}.
All these works, however, did not address the critical dynamics of an isothermal fluid.

In this work we analyze the fluctuating hydrodynamic equations of an isothermal, compressible nonideal fluid whose static properties are governed by a Ginzburg-Landau free energy functional. It is demonstrated that the isothermal condition leads to decisively different dynamic critical properties than in the standard model H universality class.
A scaling analysis of the leading self-energy contributions emerging from the nonlinear Langevin equations shows that the order-parameter dynamics effectively reduces, in the long-time limit, to a time-dependent Ginzburg-Landau model for a non-conserved order parameter, known as \textit{model A} \cite{hohenberg_halperin}.
The upper critical dimension, $d_c=4$, is the same in statics and dynamics.
The bulk viscosity diverges at the critical point with an exponent larger than in the mean-field limit, leading to overdamped sound modes at criticality. The shear viscosity, in contrast, remains finite in three dimensions, but is predicted to diverge by a power-law in two-dimensions.
The theoretical analysis is complemented by lattice Boltzmann simulations of the fluctuating hydrodynamics equations in two dimensions. We find that, even in this low dimensionality, the values of the critical exponents for the order parameter and bulk viscosity agree well with the analytical predictions that are obtained close to $d_c$ based on pure model-A behavior.

\begin{table*}[t]
\centering
\begin{tabular}{p{4.1cm}|cc|cc}
\hline\hline
   & \multicolumn{2}{c|}{isothermal} & \multicolumn{2}{c}{model H} \\
\hline 
  & \multicolumn{2}{c|}{} & \multicolumn{2}{c}{}\\[-6pt]
order parameter & \multicolumn{2}{c|}{mass density} & \multicolumn{2}{c}{entropy density}\\[-2pt]
 & \multicolumn{2}{c|}{ } & \multicolumn{2}{c}{(concentration)}\\[4pt]
o.p. relaxation mechanism& \multicolumn{2}{c|}{sound waves} & \multicolumn{2}{c}{thermal diffusion} \\[-2pt]
& \multicolumn{2}{c|}{ } & \multicolumn{2}{c}{(concentration diffusion)}\\[4pt]
relevant nonlinearity for non-classical$^1$ behavior & \multicolumn{2}{c|}{thermodynamic pressure} & \multicolumn{2}{c}{advection term}\\[16pt]
sound speed & \multicolumn{2}{c|}{isothermal, $c_s^2\sim \xi^{-\gamma/\nu}$} & \multicolumn{2}{c}{adiabatic, $c_s^2\sim \xi^{-\alpha/\nu}$ }\\[6pt]
\hline
 critical indices  & 2D & 3D & 2D & 3D \\
\hline 
  & \multicolumn{2}{c|}{} & \multicolumn{2}{c}{}\\[-6pt]
order-parameter relaxation rate, $\Gamma\propto \xi^{-z}$ & \multicolumn{2}{c|}{$z=2-\eta+x$}  & \multicolumn{2}{c}{$z=d+y$}\\[-10pt]
 & $2.2\pm 0.1$ & 2.08 $^2$ & 1.98\ldots 2.16 $^3$ & 3.07 $^4$ \\[16pt]
bulk viscosity$^5$, $\zeta_b \propto \xi^{x}$ & \multicolumn{2}{c|}{$x\sim 1.7\eta$} & \multicolumn{2}{c}{$x=z-\alpha/\nu$} \\
 & $0.45\pm 0.1$ & 0.12 $^2$ & 1.98\ldots 2.16 $^3$ & 2.9 $^6$ \\[6pt]
shear viscosity, $\zeta_s\propto \xi^{y}$ & \multicolumn{2}{c|}{$y=z-d$} & \multicolumn{2}{c}{ }\\
 & $0.2\pm 0.1$ $^7$ & finite & -0.02\ldots 0.16 $^3$ & 0.07 $^4$ \\[3pt]
\hline\hline
\end{tabular}
\caption{Comparison of characteristic properties and critical indices of an isothermal compressible fluid and an energy-conserving pure fluid (or a binary fluid at the demixing point) described by model H. The cited numerical values are rounded, see the original works for more detailed predictions. Remarks: $^1$``non-classical'' refers to deviations from predictions of van-Hove theory, which assumes constant kinetic coefficients (see text). $^2$Theoretical predictions based on model A \cite{canet_nprg_modelA_jphysA2007, krinitsyn_prudnikov_ising_2006}. $^3$Extrapolation of the $\epsilon$-expansion results \cite{siggia_prb1976, hohenberg_halperin} to 2D. $^4$see \cite{burstyn_sengers_prl1980, burstyn_sengers_pra1982, berg_moldover_prl1999, berg_moldover_pre1999, hao_viscosity_pre2005}. $^5$In the isothermal fluid, $x$ characterizes the divergence of the longitudinal [eq.~\eqref{long_visc}] rather than the bulk viscosity [see eq.~\eqref{long_visc_rg}]. Asymptotically, however, $\zeta_l\sim\zeta_b$, since the divergence of the shear viscosity is expected to be weak. $^6$see \cite{onuki_pre1997}. $^7$Prediction of the scaling theory. Present simulations could only reveal a finite critical contribution to the shear viscosity. See text for further discussion. Note: Static critical exponents have Ising values \cite{stanley_book, chaikin_book} and are identical for the isothermal and conventional (model H) fluid. $\xi$ is the correlation length, $\nu$ is the correlation length exponent, $\gamma=(2-\eta)\nu$ the susceptibility exponent, $\eta$ the anomalous dimension exponent and $\alpha$ the specific heat exponent.}
\label{tab:crit_fluid}
\end{table*}

In order to appreciate the difference of the isothermal critical dynamics from that of a non-isothermal fluid, it is useful to recapitulate the basic results of the model H universality class.
The original, incompressible model H consists of an advection-diffusion equation for the order parameter $\phi$, which is coupled to a transverse velocity field $\uv$ \cite{kawasaki_annphys1970, halperin_siggia_prl1974,siggia_prb1976, hohenberg_halperin, onuki_book},
\begin{align}
\pd_t \phi &= -\nabla\cdot(\phi\uv) + \lambda\nabla^2 \frac{\delta}{\delta \phi}\Fcal + \pi\,,\label{modelH_cont}\\
\rho_0 \pd_t \uv &= -\left(\phi\nabla \frac{\delta}{\delta \phi}\Fcal\right)_\perp + \zeta_s \nabla^2\uv + \bar\pi_\perp\,,\label{modelH_nse}
\end{align}
where the label $\perp$ indicates that the transverse projection should be taken, i.e., the projection orthogonal to the wavevector in Fourier space (cf.~sec.~\ref{sec:theor}).
In the above equations, $\rho_0$ is the mass density of the fluid, $\Fcal$ is a Ginzburg-Landau free-energy functional, $\lambda$ is a bare kinetic coefficient (e.g., thermal conductivity), $\zeta_s$ is the shear viscosity and $\pi$ and $\bar\pi$ are appropriate noise sources.
Importantly, in the case of a pure fluid at the liquid-vapor critical point, the relevant dynamical order parameter $\phi$ is the entropy density, whereas for a binary fluid at the demixing point, $\phi$ represents the concentration \cite{kawasaki_annphys1970, halperin_siggia_prl1974, siggia_prb1976, hohenberg_halperin}. 
The longitudinal part of the momentum density is neglected in the original model H, as the fluid is assumed to be incompressible, $\nabla\cdot \uv=0$.
This is a valid approximation at criticality, since thermal conduction (or, correspondingly, concentration diffusion in a binary fluid) proceeds on a much longer timescale than the propagation of sound waves. The latter process is therefore irrelevant to the dynamics of the order parameter \cite{kadanoff_swift_pr1968, kawasaki_annphys1970}.
In model H, the order-parameter relaxation rate is given in Fourier space by 
\beq \Gamma \sim \frac{\lambda k^2}{\chi(k)} \propto k^z\,,
\label{th_relax}
\eeq
with $\chi(k)$ being the susceptibility (isothermal compressibility) \cite{hohenberg_halperin, onuki_book}. This relation can be derived from eqs.~\eqref{modelH_cont} and \eqref{modelH_nse} by linearizing and neglecting the advection term.
Since $\chi(k)\propto k^{-2+\eta}$, the ``classical'' (van Hove \cite{vanHove_physrev1954}) theory predicts a dynamic critical exponent of $z=4-\eta$ for model H.
The classical result, however, turns out to be violated in a real fluid, since the kinetic or transport coefficients are, due to the presence of reversible mode-couplings, affected by the critical order-parameter fluctuations as well \cite{hohenberg_halperin, kawasaki_review1976, mazenko_book, bhattacharjee_book}.
Specifically, in model H, the kinetic coefficient $\lambda$ is renormalized by the advective coupling between $\phi$ and $\uv$, changing the dynamical exponent to $z=4-\eta-z_\lambda=d+y$, where $z_\lambda$ and $y$ are the exponents characterizing the divergence of $\lambda$ and the shear viscosity $\zeta_s$, respectively \cite{halperin_siggia_prl1974, siggia_prb1976, kawasaki_review1976, ohta_kawasaki_ptp1976, burstyn_sengers_prl1980, burstyn_sengers_pra1982}.
The effect of critical fluctuations on the shear viscosity, $\zeta_s \propto \xi^y$, is weak, leading only to a small exponent of $y\simeq 0.07$ in 3D \cite{siggia_prb1976, kawasaki_review1976, ohta_kawasaki_ptp1976, hao_viscosity_pre2005}, as confirmed by experiments \cite{berg_moldover_prl1999, berg_moldover_pre1999}. 

Critical fluctuations in a pure fluid have an effect on sound waves as well, which, however, is one-sided since the latter are decoupled from the order-parameter dynamics. Acoustic effects can be studied with an extended, compressible version of model H that includes the full set of equations for the mass, momentum and energy density \cite{kroll_ruhland_physlett1980, kroll_ruhland_pra1981, dengler_schwabl_epl1987, dengler_schwabl_zphysb1987, folk_moser_jlowtemp1995, folk_moser1_pre1998, folk_moser2_pre1998, folk_moser3_pre1999, adzhemyan_jetp1998, onuki_pre1997} (see also \cite{bhattacharjee_sound_rpp2010} and references therein).
In an energy-conserving pure fluid, sound waves propagate with the \textit{adiabatic} speed of sound \cite{Landau_FluidMech59, Hansen_ThoSL, BoonYip_book}, 
\beq c_{s,\text{ad}}^2 = \frac{\pd p}{\pd \rho}\Big|_S = \frac{c_p}{c_V} \frac{1}{\rho\chi}\,,
\label{adiab_sound}
\eeq
where $p$ is the pressure and $c_p$, $c_V$ are the specific heats at constant pressure and volume. These are related by $c_p = c_V + T\chi \beta_V^2/\rho$, where $\beta_V=(\pd p / \pd T)|_\rho$ is the thermal pressure coefficient (slope of the $p$-$T$-curve). Since $\beta_V$ is not critical, we have $c_p \sim \chi$ and hence the critical behavior of the speed of sound (at zero frequency) is given by \cite{folk_moser1_pre1998, onuki_pre1997,onuki_book, bhattacharjee_sound_rpp2010}
\beq c_{s,\text{ad}}^2\sim c_V^{-1} \propto \xi^{-\alpha/\nu}\,. 
\eeq
For comparison, in a binary fluid, the critical sound speed is governed by the constant-pressure specific heat, whose divergence is -- due to a larger background contribution -- much less pronounced than for the pure fluid.
The critical sound damping is given by the bulk viscosity, which can be determined from a Green-Kubo relation involving the nonlinear pressure fluctuations \cite{kroll_ruhland_physlett1980, kroll_ruhland_pra1981, dengler_schwabl_epl1987, onuki_pre1997}.
The extended model H predicts a strongly diverging bulk viscosity, $\zeta_b \propto \xi^x$ (at zero frequency), with $x=z-\alpha/\nu$ being $\simeq 2.9$ in 3D.

While theoretical and experimental investigations of critical dynamics in pure or binary fluids have a long history, simulations seem to be scarce and have been performed only quite recently \cite{das_binary_prl2006, das_binary_jcp2006, jagannathan_prl2004, chen_dynamic_prl2005, hamanaka_onuki_pre2005, roy_das_epl2011}.
However, most of these studies are in 3D, and values of the dynamic critical exponents for model H in 2D seem so far not to have been obtained either by experiment or simulations (cf.~\cite{casalnuovo_prl1982, casalnuovo_pra1984, calvo_ferrel_pra1985, calvo_pra1985, bhattacharjee_restrict_prl1996, zhang_shorttime_2000, honerkamp_smith_prl2012}). The values given in Tab.~\ref{tab:crit_fluid} for model H in 2D therefore represent extrapolations of theoretical $\epsilon$-expansion results \cite{siggia_prb1976, hohenberg_halperin} (throughout this paper $\epsilon=4-d$). The shear-viscosity exponent $y$ in Tab.~\ref{tab:crit_fluid} has been obtained using the $\epsilon$-expansion result for $z_\lambda$ [$z_\lambda=(18/19)\epsilon(1-0.003\epsilon)+O(\epsilon^3)$]  in conjunction with the relation $y=4-d-\eta-z_\lambda$, giving the lower bound, as well as the direct $\epsilon$-expansion result for $y$ [$y=(1/19)\epsilon(1+0.238\epsilon)+O(\epsilon^3)$], giving the upper bound. Since the $O(\epsilon^2)$-term in the $\epsilon$-expansion of $z_\lambda$ is quite small relative to the leading term, one might suspect that the extrapolated value will not be grossly unrealistic.
Of course, one has to keep in mind that, transport coefficients in a two-dimensional fluid generally acquire logarithmic divergences in the long-time or -wavelength limit \cite{pomeau_pra1972, pomeau_mct_review1975, forster_nelson_prl1976, forster_nelson_pra1977}, which might interfere with possible critical divergences. 

From relation \eqref{adiab_sound} we see that a description in terms of the isothermal speed of sound, 
\beq c_{s,\text{iso}}^2= \frac{\pd p}{\pd \rho}\Big|_T=\frac{1}{\rho \chi}\,,
\eeq 
would become applicable if the specific heat ratio $c_p/c_V$ would be close to 1. 
Dynamically, adiabatic conditions are achieved if the thermal relaxation rate [eq.~\eqref{th_relax}] is much smaller than the characteristic frequency of a sound wave, i.e., 
\beq \Gamma \ll c_s k\,.
\label{adiab_cond}\eeq 
Since $c_{s,\text{ad}} k \propto \xi^{-\alpha/2\nu-1}$ for $k\sim \xi^{-1}$, the above relation is clearly fulfilled in a ordinary fluid close to the critical point. Relation \eqref{adiab_cond} provides \textit{a posteriori} also a justification for neglecting the ``faster'' acoustic processes in the usual model H calculations. Far from criticality, violations of condition \eqref{adiab_cond} can occur at finite wavenumbers in the hydrodynamic regime \cite{bencivenga_sound_epl2006}.
For small wavenumbers, isothermal conditions can be achieved by coupling to fluid to some kind of heat bath, so that temperature fluctuations are removed at a sufficiently fast rate. At the same time, the friction between fluid and substrate must be kept sufficiently small, in order not to break momentum conservation and violate the characteristic sound mode behavior of the compressible fluid in the relevant wavenumber regime \cite{ramaswamy_mazenko_pra1982}.

After these introductory remarks on the critical dynamics of ordinary non-ideal fluids, we now turn to the analysis of the isothermal non-ideal fluid.

\section{Theory}
\label{sec:theor}
An isothermal compressible fluid is governed by a continuity equation for the mass density $\rho$ and a conservation equation for the momentum density $\jv\equiv \rho\uv$ \cite{Landau_FluidMech59, felderhoff_physica1970, langer_turski_pra1973, onuki_book},
\beq \pd_t \rho = -\nabla \cdot \jv\,, \label{cont1}\eeq
\beq \pd_t \jv =  -\rho\nabla\frac{\delta \Fcal}{\delta\rho} + \zeta_s \nabla^2 \frac{\jv}{\rho} + (\zeta_b + \zeta_s[1-2/d])\nabla\nabla\cdot \frac{\jv}{\rho} + \nabla\cdot \ten R \,. \label{nse1}\eeq
Here, $\zeta_s$ and $\zeta_b$ are the bare shear and bulk viscosity and $\ten R$ is a random stress tensor with correlations \cite{Landau_FluidMech59, KimMazenko_1991}
\beq \bra R_{\alpha \beta}(\bv{r},t) R_{\gamma \delta}(\bv{r'},t') \ket = 2k_B T \left[ \zeta_s \left(\delta_{\alpha \gamma} \delta_{\beta \delta} + \delta_{\alpha \delta} \delta_{\beta \gamma} - \frac{2}{d} \delta_{\alpha \beta}\delta_{\gamma \delta}\right)
+ \zeta_b\, \delta_{\alpha \beta}\delta_{\gamma \delta}\right]\delta(\bv{r}-\bv{r'})\delta(t-t')
\,,
\label{rand_stress}\eeq
imparting Gaussian thermal noise on the fluid.
The static probability distribution of the density fluctuations are governed by the Ginzburg-Landau free energy functional
\beq 
\Fcal=\int d\rv \left[\onehalf \kappa (\nabla \phi)^2 + f_0(\phi)\right]\,,
\label{fef}
\eeq 
where $\kappa$ is a constant and 
\beq \phi\equiv (\rho-\rho_0)/\rho_0=\delta\rho/\rho_0
\eeq 
is the order parameter. $\rho_0$ is a constant background density. The Landau free energy density $f_0$ is taken to be a quartic polynomial in $\phi$,
\beq f_0(\phi) = \onehalf r \phi^2 + \frac{1}{4}u \phi^4\,,
\eeq
where $r$ and $u$ are free parameters.
The ``streaming term'' involving $\Fcal$ in eq.~\eqref{nse1} can be written as a divergence of a pressure tensor $\ten P$ \cite{felderhoff_physica1970, yang_molecular_1976, evans_interface_1979, lowengrub_1988, jasnow_vinals_1996, anderson_diffuse_1998},
\beq \rho\nabla\frac{\delta \Fcal}{\delta\rho} = \nabla\cdot \ten P = \nabla p_0 - \kappa' \rho \nabla\nabla^2 \rho
\eeq
where $\kappa'\equiv \kappa/\rho_0^2$,
\beq P_{\alpha\beta} = \left(p_0 - \kappa' \rho \nabla^2 \rho - \frac{\kappa'}{2}|\nabla\rho|^2\right) \delta_{\alpha\beta} + \kappa' (\pd_\alpha \rho) (\pd_\beta \rho)\,,
\label{press_ten}
\eeq
and $p_0$ is a scalar pressure given by
\beq \begin{split}
p_0 &= \rho \pd_\rho f_0  - f_0 = r \phi + \onehalf r \phi^2 + u \phi^3 + \frac{3}{4} u \phi^4\,.
\end{split}
\label{p0_nlin}\eeq

An essential complication in the analysis of the compressible Navier-Stokes equations is the presence of the nonlinearity $\jv/\rho$ in the viscous stress \cite{das_mazenko_pra1986, kawasaki_mct_review_1995, das_book}. 
Here, we treat this term perturbatively by expanding $1/\rho$ around the background density $\rho_0$, i.e.\ $1/\rho = 1/\rho_0 - (1/\rho_0^2) \delta\rho + \ldots$ \cite{mazenko_yeo_jstatp1994}. We will show below that this term is irrelevant for the critical behavior of the fluid above two dimensions.
Furthermore, the convection term $\nabla(\jv\jv/\rho)$ has been omitted in the above Navier-Stokes equations. The effect of this term has been studied extensively in the incompressible case \cite{pomeau_pra1972, pomeau_mct_review1975, forster_nelson_prl1976, forster_nelson_pra1977} and is known to renormalize the shear and bulk viscosity by a finite amount above two dimensions and by a logarithmically divergent contribution in 2D.
In principle, this term requires careful treatment also in the case of a compressible fluid (cf.~\cite{staroselsky_orszag_prl1990}), taking into account a possible interplay with critical fluctuations. This, however, is out of the scope of the present work, and in the subsequent analysis, it is therefore assumed that the ensuing renormalizations have already been performed on the bare quantities or can be considered separately from critical fluctuations. 

To proceed, eqs.~\eqref{cont1} and \eqref{nse1} are written in Fourier space,
\beq \omega \drho = \kv\cdot \jv\,,
\label{cont2}
\eeq
\beq -\im \omega \jv = -\im \kv c_s^2(\kv)\drho - \im \kv p\st{nl} - \im \bv{N} - \nu_s k^2\left[ \jv+\bv{Y}\right] - (\nu_b + \nu_s[1-2/d]) \kv \kv\cdot \left[\jv+\bv{Y}\right] + \im\kv\cdot \ten R + \im \kv\cdot \hv \,,
\label{nse2}
\eeq
where $\nu_s=\zeta_s/\rho_0$ and $\nu_b=\zeta_b/\rho_0$ are the kinematic shear and bulk viscosities. The generalized isothermal speed of sound \footnote{This expression for the speed of sound holds only in the supercritical regime. Below the critical point ($r<0$), the nonlinear terms provide an additional contribution, so that $c_s^2$ is positive in each bulk phase.
}
\beq c_s^2(\kv) = c_s^2 + \kappa'\rho_0 k^2 = (r+\kappa k^2)/\rho_0\,,
\label{gen_sound_r}
\eeq
contains the linear part of the thermodynamic pressure, while $p\st{nl}$ and $\bv{N}$ are the Fourier-transforms of the remaining nonlinear parts:
\beq p\st{nl}(\kv,\omega) = \frac{1}{2}r \int_{\tilde q} \phi(\tilde k-\tilde q)\phi(\tilde q) + u \int_{\qt,\qt'} \phi(\kt-\qt-\qt') \phi(\qt) \phi(\qt') + \frac{3}{4}u \int_{\qt,\qt',\qt''} \phi(\kt-\qt-\qt'-\qt'') \phi(\qt)\phi(\qt')\phi(\qt'')\,,
\label{pnl}
\eeq
\beq \bv{N}(\kv,\omega) = \kappa \int_{\qt} \phi(\kt-\qt) \qv q^2 \phi(\qt) \,.
\label{kappa_nl}\eeq
Here, the shorthand notation $\kt\equiv (\kv,\omega)$, $\qt\equiv (\qv,\sigma)$, etc.\ and $\int_{\qt} \equiv \int \frac{d \qv}{(2\pi)^d} \frac{d \sigma}{(2\pi)}$ is introduced.
The quantity 
\beq \bv{Y}(\kv,\omega) = \int_{\qt} \jv(\qt)\phi(\kt-\qt)
\label{j_nonlin}
\eeq
represents the leading correction term of the expansion of $\jv/\rho$ in the viscous stress around $\jv/\rho_0$.

The nonlinear Navier-Stokes equations \eqref{cont2} and \eqref{nse2} can be split into longitudinal and transverse parts with respect to the wavevector $\kv$. The corresponding longitudinal (`l') and transverse (`t') projections of a vectorial quantity $\vv = v_l \hat \kv + \vv_t$ are defined as $v_l = \hat \kv \cdot \vv$ and $\vv_t = \mathcal{T}_\kv\vv$, where $\hat \kv\equiv \kv/k$ and $\mathcal{T}_\kv \equiv (\Imat - \hat \kv \hat \kv)$. Analogously, for a tensorial quantity like $\ten R$, we have $\hat \kv\cdot \ten R = R_l\hat \kv + \bv{R}_t$, with $R_l \equiv \hat \kv\cdot \ten R \cdot \hat \kv$, and $\bv{R}_t \equiv \hat \kv\cdot \ten{R} \cdot \mathcal{T}_\kv$. 
We thus arrive at
\beq \omega \drho = k j_l \,,\label{contl}\eeq
\beq \omega j_l = k c_s^2(\kv) \drho + k p\st{nl} + N_l - \im \nu_l k^2 (j_l+Y_l) - k R_l - k h_l \,, \label{nsel}\eeq
\beq \omega \jv_t = \bv{N}_t - \im\nu_t k^2 (\jv_t+\bv{Y}_t) - k \bv{R}_t - k \bv{h}_t \,,\label{nset}\eeq
where 
\beq \nu_l = \nu_b + \nu_s(2-2/d)
\label{long_visc}
\eeq
denotes the longitudinal and $\nu_t = \nu_s$ the transverse viscosity.
The longitudinal and transverse parts of the random stress tensor are correlated as $\bra |R_l(\kv,\omega)|^2\ket = 2 \rho_0 \nu_l k_B T $ and $\bra |R_{t,\alpha}(\kv,\omega)|^2\ket = 2 \rho_0 \nu_t k_B T $.
Combining eqs.~\eqref{contl} and \eqref{nsel}, the longitudinal current $j_l$ can be eliminated completely, leaving only eq.~\eqref{nset} for the transverse current and a single, nonlinear sound-wave equation for the order parameter (setting henceforth $\rho_0=1$): 
\beq -\omega^2 \phi + k^2 c_s^2(\kv)\phi - \im \omega \nu_l k^2\phi = -k^2 p\st{nl} - k N_l + \im \nu_l k^3 Y_l + k^2 R_l + k^2 h_l \,.\label{crit_den_eq}\eeq
Obviously, the term $\bv Y$, which can be written as
\beq \bv Y = \int_{\qt} \left[ \phi(\qt)\frac{\sigma}{q}\hat \qv  + \jv_t(\qt)\right] \phi(\kt-\qt)\,,\eeq
provides a bi-directional coupling between the order parameter and the transverse current. Additionally, the transverse current is affected by the order-parameter fluctuations through the term $\bv N_t$.
We remark that, due to the way the free energy functional enters the Navier-Stokes equations, there appear to be more nonlinearities in eq.~\eqref{crit_den_eq} than in the corresponding static critical theory or in nonlinear sound-wave equations studied in isotropic elastic phase transitions \cite{folk_schwabl_elastic_prb1979, schwabl_sound_jstatp1985}. In particular, in the latter, the restoring force is given by a term of the form $\delta \mathcal{F} / \delta\rho$ rather than by a pressure gradient.

\subsection{Linear hydrodynamics}
\begin{figure}[t]\centering
   (a)\includegraphics[width=0.3\linewidth]{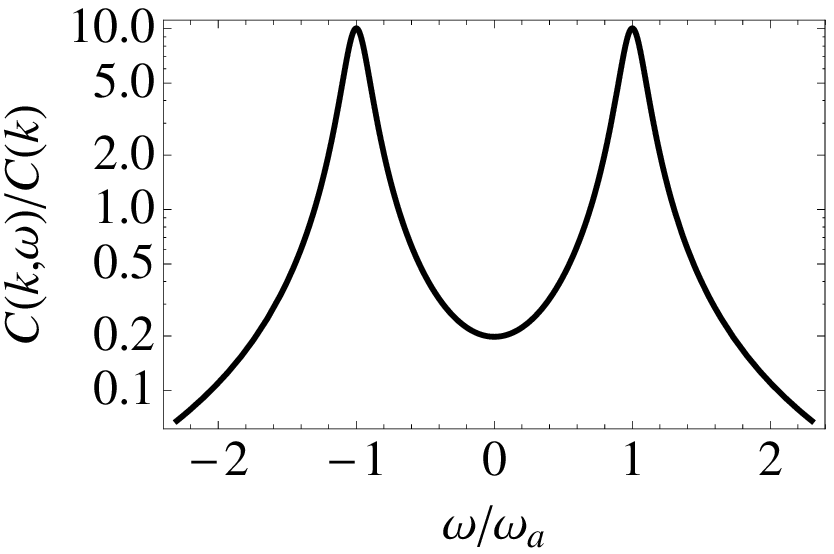}
   (b)\includegraphics[width=0.3\linewidth]{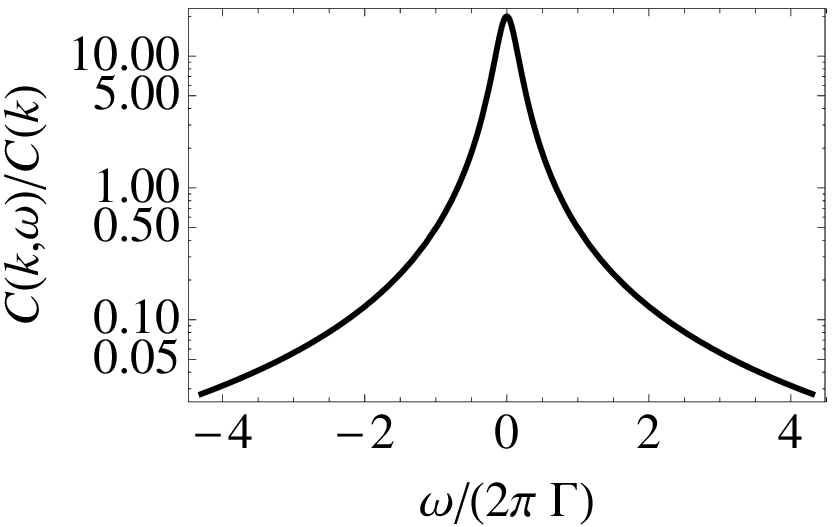}
   (c)\includegraphics[width=0.3\linewidth]{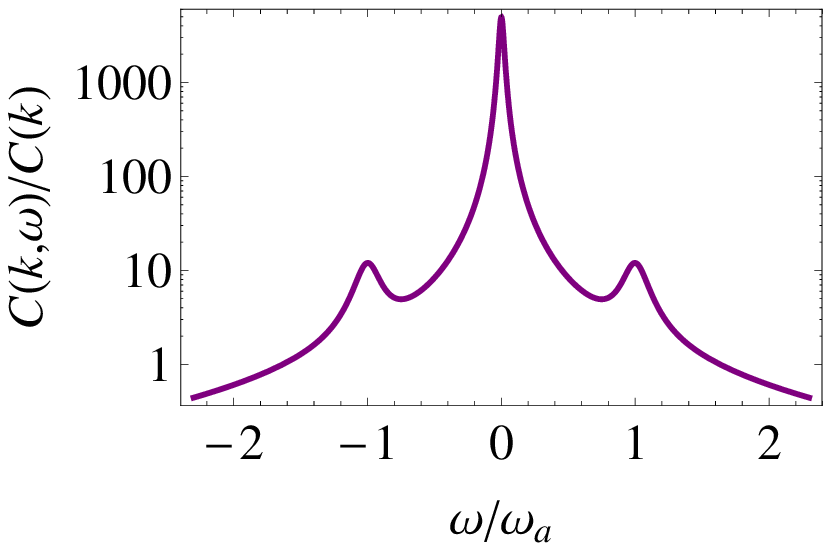}
   \caption{Typical shape of the dynamic structure factor $C(k,\omega)$ of an isothermal fluid in the weak-damping (a) and strong-damping (b) regime. The latter case is realized at the critical point (see text). For comparison, (c) shows the structure factor of an ordinary, energy-conserving fluid close to the critical point.}
    \label{fig:sfdyn_peaks}
\end{figure}

At this point it is useful to collect some basic results of the linearized model.
Neglecting the nonlinear terms in eq.~\eqref{crit_den_eq}, the \textit{bare} response- and correlation function (labeled by the index 0) for the order parameter are given by
\beq G_0(\kv,\omega) \equiv \frac{\delta \bra \phi(\kv,\omega) \ket}{\delta h_l(\kv,\omega)}= \frac{k^2}{-\omega^2 + k^2 c_s^2(\kv) - \im \omega\nu_l k^2}\,,
\label{G0}
\eeq
\beq
C_0(\kv,\omega) \equiv \frac{\bra \phi(\kv,\omega) \phi(\kv',\omega')\ket }{ (2\pi)^{d+1} \delta(\kv+\kv')\delta(\omega+\omega')}= \frac{2\nu_l  k_B T k^4}{\big[\omega^2  -k^2 c_s^2(\kv) \big]^2 + (\omega \nu_l  k^2)^2}\,.
\label{C0}
\eeq
The response and correlation functions are related by a fluctuation-dissipation theorem,
\beq C_0(\kv,\omega) = \frac{2 k_B T}{\omega} \IM G_0(\kv,\omega)\,,
\label{fdt_lin}
\eeq
which, in the zero-frequency limit, becomes the fluctuation-response relation
\beq  C_0(\kv) = k_B T G_0(\kv,\omega=0),
\label{fluct_resp}
\eeq
where $C_0(\kv)$ is the static structure factor,
\beq C_0(\kv) = \int \frac{d\omega}{2\pi} C_0(\kv,\omega) = \frac{k_B T}{c_s^2(\kv)} \,. \label{sf_stat}\eeq
In the mean-field limit, the susceptibility is given by $\chi_0=1/r$ for $r>0$ and $\chi_0=-1/2r$ for $r<0$, and is related to the speed of sound by $c_s^2=1/\rho_0 \chi_0$. 
The correlation length for purely Gaussian fluctuations is given by $\xi_0=\sqrt{\kappa \chi_0}$.

In the linearized case, eq.~\eqref{crit_den_eq} represents a damped harmonic oscillator driven by random noise \cite{chaikin_book}. The dispersion relation of the associated sound waves is given by
\beq \omega = \pm \left[ c_s^2(k) k^2 - \nu_l^2 k^4/4\right]^{1/2} - \im \nu_l k^2 / 2\,.
\label{sound_disp}
\eeq
In the weakly damped case, where $4c_s^2(k) >\nu_l^2 k^2$, sound waves are oscillating with frequencies $\omega_a= \pm k \left[ c_s^2(k) - \nu_l^2 k^2/4\right]^{1/2}\approx c_s(k)k$ and are exponentially damped with a rate of $\nu_l k^2/2$. In the opposite, overdamped case, the solution \eqref{sound_disp} becomes purely imaginary and sound waves decay, in the limit of long times, with a rate of 
\beq \Gamma(\kv) = \frac{c_{s}^2(\kv)}{\nu_{l}}\,.
\label{sound_overdamp}
\eeq
At short times, another decay regime characterized by a rate $\sim \nu_l k^2$ is present. This regime is negligible for strong damping, that is, for $c_s^2\ll \nu_l^2 k^2$.
The response and correlation functions in the overdamped, long-time limit can be simply obtained by neglecting the ``inertial'' term $\omega^2$ in eqs.~\eqref{G0} and \eqref{C0}, yielding
\beq G_0(\kv,\omega) = \frac{1}{- \im \omega\nu_{l} +c_{s}^2(\kv)}\,, \label{dyn_resp_od}
\eeq
\beq 
C_0(\kv,\omega) = \frac{1}{\nu_{l}}\frac{2 k_B T}{\omega^2 + \Gamma^2(\kv)}\label{dyn_correl_od}\,.
\eeq
In the time-domain, this corresponds to a pure exponential decay:
\beq G_0(\kv,t) = \frac{1}{\nu_{l}}\exp\left[- \Gamma(\kv) t\right] \theta(t), \label{dyn_resp_time}\eeq
\beq C_0(\kv,t) = C_0(\kv) \exp[-\Gamma(\kv) |t|]\,.\label{dyn_correl_time}\eeq
Fig.~\ref{fig:sfdyn_peaks}a,b show the typical shape of the linear dynamic structure factor of the isothermal fluid in the weak and strong damping case. We will see that, close to criticality, long-wavelength order-parameter fluctuations in the isothermal fluid are always overdamped, causing the two sound-mode peaks in Fig.~\ref{fig:sfdyn_peaks}a to merge to a single peak located at zero frequency (Fig.~\ref{fig:sfdyn_peaks}b). For comparison, the dynamic structure factor of an ordinary fluid  (Fig.~\ref{fig:sfdyn_peaks}c) is characterized by two sound-mode peaks at finite frequencies and a central peak originating from thermal diffusion, which dominates the total intensity close to the critical point. 

Similarly, for the transverse current, we obtain from eq.~\eqref{nset} the bare response and correlation functions
\beq G_{t,0}(\kv,\omega) = \frac{k}{\omega + \im\nu_t k^2}\,,
\label{Gt0}
\eeq
\beq C_{t,0}(\kv,\omega) = \frac{2\nu_t k_B T k^2}{\omega^2 + (\nu_t k^2)^2}\,,
\label{Ct0}
\eeq
where $\bra j_{t,\alpha}(\kv,\omega) j_{t,\beta}(\kv',\omega')\ket = C_{t,0}(\kv,\omega) (2\pi)^{d+1}\delta(\kv+\kv')\delta(\omega+\omega')\delta_{\alpha\beta}$.
The static correlations of the transverse current are independent of the wavenumber,
\beq C_{t,0}(\kv) =  k_B T\,.\label{Ct_static}
\eeq

The linear hydrodynamics expressions \eqref{G0} and \eqref{C0} can be cast into standard dynamical scaling forms \cite{hohenberg_halperin, ma_book, chaikin_book}, 
\beq G_0(\kv,\omega) = \xi^{2-\eta} \mathcal{G}(k\xi,\omega \xi^z),\quad 
C_0(\kv,\omega) = \xi^{2-\eta+z} \mathcal{C}(k\xi,\omega \xi^z)\,,
\label{C_scalform}
\eeq
with $\mathcal{G}$ and $\mathcal{C}$ being scaling functions, $\eta=0$ and $z=2$ a dynamic scaling exponent. This value of $z$ can also be directly inferred from the damping rate in the overdamped case, eq.~\eqref{sound_overdamp}.
Analogously, for the transverse current we have from eqs.~\eqref{Gt0} and \eqref{Ct0}:
\beq G_{t,0}(\kv,\omega) = \xi^{z_t-1} \mathcal{G}_t(k\xi,\omega \xi^{z_t}),\quad 
C_{t,0}(\kv,\omega) = \xi^{z_t}\mathcal{C}_t(k\xi, \omega \xi^{z_t})\,,
\label{Ct_scalform}
\eeq
with a dynamic exponent of $z_t=2$.

\subsection{Critical order-parameter dynamics}
\begin{figure}[t]
\centering
\includegraphics[width=0.95\linewidth]{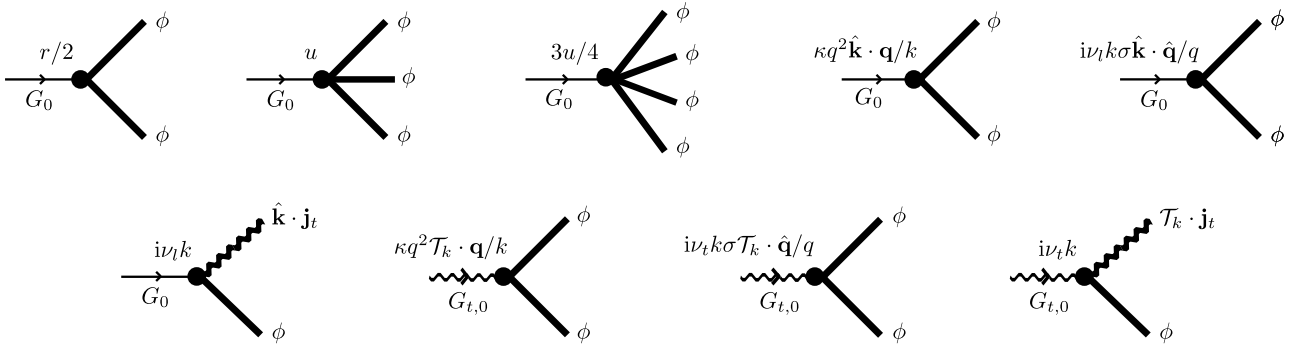}
\caption{Fundamental vertices of the model arising from eqs.~\eqref{pnl}, \eqref{kappa_nl} and \eqref{j_nonlin}. Here, $\kv$ is an external wavevector, while $\qv$ and $\sigma$ denote an internal wavevector and frequency. A filled circle represents a coupling constant and an integration over internal wavevectors and frequencies respecting space- and time-translational invariance.}
\label{fig:feyn}
\end{figure}

\begin{figure*}[t]
\centering
\includegraphics[width=0.98\linewidth]{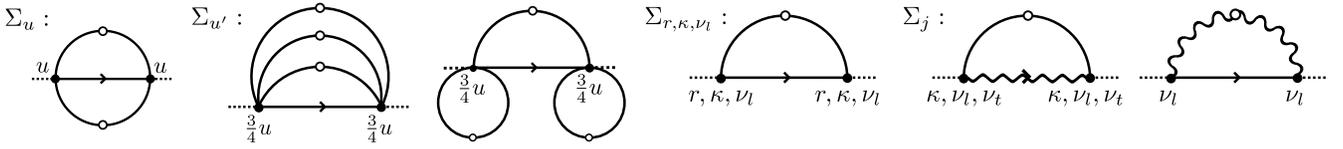}\quad 
\caption{Leading frequency-dependent diagrams contributing to the self-energy of the order parameter, $\Sigma(\kv,\omega)$. 
A solid (wavy) line with an arrow represents $G_0$ ($G_{t,0}$) and a solid (wavy) line with an open circle $C_0$ ($C_{t,0}$). Dashed lines indicate amputated external ``legs'' for clarity. Note that only certain combinations of couplings are admissible in the diagrams of $\Sigma_j$ (cf.\ Fig.~\ref{fig:feyn}).}
\label{fig:self_en_dyn}
\end{figure*}

The critical dynamics of the order parameter, governed by eq.~\eqref{crit_den_eq}, is discussed here within a mode-coupling approach \cite{kawasaki_annphys1970, kawasaki_review1976, kawasaki_gunton_mcrg_prb1976, luettmer_strathmann_sengers_jcp1995, onuki_book, bhattacharjee_book}. To this end, we construct a perturbative solution of the nonlinear order-parameter equation using the response function formalism \cite{ma_mazenko_prb1975, ma_book, ferrel_eqnmotion_1979, mazenko_book} and identify the leading contributions via a scaling analysis.
With the help of the bare response function $G_0$, eq.~\eqref{crit_den_eq} can be rearranged as
\beq \phi= \phi_0 - G_0 p\st{nl} - G_0 N_l/k + \im \nu_l k G_0 Y_l + G_0 h_l\,,
\label{den_pert}\eeq
with $\phi_0 = G_0 R_l \,\label{den_sol0}$
being the zeroth-order solution.
Due to the coupling between the order parameter and the transverse current, we also need to consider eq.~\eqref{nset}, which can be written as
\beq \jv_t = -\jv_{t,0} + G_{t,0} \bv{N}_t/k - \im \nu_t k \bv{Y}_t - G_{t,0}\bv{h}_t\,,
\label{jt_pert}
\eeq
with $\jv_{t,0} = G_{t,0}\bv{R}_t$.

The nonlinearities on the right hand side of eqs.~\eqref{den_pert} and \eqref{jt_pert} can be translated into the diagrammatic representation given by Fig.~\ref{fig:feyn}. There, a solid (wavy) line with an arrow represents $G_0$ ($G_{t,0}$), a thick solid (wavy) line represents the order parameter $\phi$ (the transverse current $\jv_{t,0}$), a filled circle denotes a coupling constant and an integration over internal wavevectors and frequencies respecting space- and time-translational invariance. 
The vertices involving the couplings $r$, $\kappa$, $\nu_l$ and $\nu_t$ have no counterparts in the static theory or in standard Ginzburg-Landau models \cite{hohenberg_halperin}. They are specific to the compressible fluid and are, for instance, known to be important in the case of a supercooled liquid \cite{das_mazenko_pra1986,das_book}.

A solution of eq.~\eqref{den_pert} for $\phi$ can be iteratively constructed, following standard rules \cite{kawasaki_annphys1970, ma_mazenko_prb1975, ma_book, mazenko_book}, and leads to a Dyson relation for the full response function $G\equiv \delta \bra \phi \ket / \delta h_l$:
\beq G(\kv,\omega) = \frac{1}{G_0^{-1}(\kv,\omega) - \Sigma(\kv,\omega)} = \frac{k^2}{-\omega^2 + k^2 c_s^2(\kv) - \im \omega\nu_l k^2 - k^2 \Sigma(\kv,\omega)}\,.
\label{gfull_selfen}
\eeq
Here, $\Sigma$ is a self-energy, which encapsulates the effect of the nonlinear interactions between the order-parameter fluctuations.
These can be understood to lead to a renormalization of the transport coefficients of the fluid of the form
\begin{align}
c_{sR}^2(\kv) &= c_s^2(\kv) - \RE\Sigma(\kv,0)\,,\label{renorm_cs}\\
\nu_{lR}(\kv) &= \nu_l +  \Pi(\kv,\omega\ra 0)\,,
\label{renorm_visc}
\end{align}
where $\Pi(\kv,\omega)\equiv  \pd \IM \Sigma(\kv,\omega)/\pd\omega$. 
We shall focus here only on the small-frequency limit and neglect any frequency-dependence of the renormalized quantities $c_{sR}$ and $\nu_{lR}$. 
Thus, we assume that the response function keeps the same principle form as in linear hydrodynamics, but with appropriately renormalized, wavenumber-dependent transport coefficients.
Also, a possible renormalization of the background density is neglected here.

The nonlinearities in the model give rise to a large number of diagrams contributing to the static and dynamic (i.e., frequency-independent and -dependent) parts of the self-energy.
The dominant contribution can in principle be determined through a straightforward scaling analysis (cf.\ \cite{kawasaki_review1976, luettmer_strathmann_sengers_jcp1995}), making use of the dynamic scaling forms of the response and correlation functions stated above.
Regarding the static parts, however, we can also directly invoke the fact that the nonlinear Langevin equations of the model preserve the equilibrium probability distribution of the Ginzburg-Landau free-energy functional \cite{KimMazenko_1991, onuki_book}. Thus, in the hydrodynamic limit, the renormalization of the isothermal speed of sound must be consistent with the static theory, implying that, 
\beq c_{sR}^2=1/\rho\chi \propto \xi^{-\gamma/\nu}\,,\eeq
where $\chi$ is the isothermal compressibility, $\xi$ the correlation length and $\gamma$ and $\nu$ are the usual static critical exponents [note $\gamma=(2-\eta)\nu$] \cite{stanley_book, sengers_supercritical_1994, onuki_book} \footnote{More generally, one would expect that $c^2_{sR}(\kv)=k_B T / C(\kv)$ for arbitrary $k$, as a consequence of a fluctuation-response relation like eq.~\eqref{fluct_resp}. Due to the presence of the $1/\rho$-nonlinearity in the equations of motion, however, there is no simple FDT analogous to eq.~\eqref{fdt_lin} connecting the full correlation and response function in the compressible fluid, except in the small-$k$ limit \cite{das_mazenko_pra1986, das_book}. An FDT can be proven for the nonlinear oscillator eq.~\eqref{crit_den_eq} if the $1/\rho$-nonlinearity is neglected, see \cite{folk_schwabl_elastic_prb1979}}. As is well known, the upper critical dimension for the present static model is $4$.
For comparison, in a conventional fluid (model H), sound waves propagate with the adiabatic speed of sound, which vanishes much more weakly at the critical point, $c^2_{sR}\propto \xi^{-\alpha/\nu}$, with $\alpha$ being the specific heat exponent \cite{stanley_book, kroll_ruhland_physlett1980,folk_moser1_pre1998, onuki_pre1997, onuki_book}.

Turning to the dynamics, the leading irreducible, frequency-dependent diagrams contributing to $\Sigma$ are shown in Fig.~\ref{fig:self_en_dyn}. 
For a given diagram $\Sigma_{(i)}$, we have $\Pi_{(i)} \sim \Sigma_{(i)} \xi^{z}$ as far as the scaling behavior is concerned. 
We do not consider here an expansion in the number of loops, but rather focus only on the leading diagrams arising from each vertex. 
The analytic expressions of the individual diagrams are given by
\begin{align}
\Sigma_{u}(\kv,\omega) &= 18 u^2  \int_{\qt,\qt'} G_0(\kt-\qt-\qt') C_0(\qt) C_0(\qt')\,,\label{self_u2}\\
\Sigma_{u'}(\kv,\omega) &= 54  u^2 \int_{\qt,\qt',\qt''} G_0(\kt-\qt-\qt'-\qt'') C_0(\qt)C_0(\qt')C_0(\qt'')\nonumber \\
& \quad + 81 u^2 \int_{\qt} G_0(\kt-\qt) C_0(\qt) \left[\int_{\qt} C_0(\qt)\right]^2\,,\\
\Sigma_{r}(\kv,\omega) &= r^2 \int_{\qt} G_0(\kt-\qt)C_0(\qt)\,,\label{self_r2}\\
\Sigma_{\kappa}(\kv,\omega) &= \kappa^2\int_{\qt} G_0(\kt-\qt)C_0(\qt)\left[(\kv\cdot\qv)^2 + O(k^4)\right]\,,\label{self_kappa}\\
\Sigma_{\nu_l}(\kv,\omega) &= - \nu_l^2 \int_{\qt} G_0(\kt-\qt)C_0(\qt) \left[\omega \sigma (\hat \kv\cdot \hat \qv)^2 + O(\omega^2)\right]\,,\\
\Sigma_{j}(\kv,\omega) &= \nu_l\nu_t \int_{\qt} G_{t,0}(\qt) C_0(\kt-\qt)\, \left[ \omega q  \left(\hat\kv\cdot \mathcal{T}_\qv \cdot \hat\kv\right) +O(\omega k)\right] \nonumber \\
& \quad \, -\nu_l^2 \int_{\qt} G_0(\qt) C_{t,0}(\kt-\qt)\left[\kv\cdot\qv + O(k^2)\right] +\ldots \label{self_j}\,.
\end{align}
In eqs.~\eqref{self_kappa} to \eqref{self_j}, only the principle form of the kernels is indicated, which is sufficient to derive scaling properties.
Also, expressions for the remaining one-loop diagrams of Fig.~\ref{fig:self_en_dyn} that involve two different couplings or a transverse current response/correlation function are not stated explicitly but can be easily obtained. 
In fact, it will not be necessary to compute them explicitly, since all vertices involving $r$, $\kappa$, $\nu_{l}$ or $\nu_{t}$ scale in the same way, up to differences of $O(\eta)$. To see this, note that $j_t$ scales like $\phi\xi^{-1}$, as can be inferred from the form of the corresponding correlation functions [eqs.~\eqref{C_scalform}, \eqref{Ct_scalform}].
Some of the additional diagrams at two-loop order are briefly discussed in appendix~\ref{app:add_diagr}; they will give rise to subdominant contributions and can thus be safely neglected.

First of all, we consider the mean-field approximation, where the values of the dynamic scaling exponents are given by $z=z_t=2$. Taking into account the strong temperature-dependence of the Landau parameter, $r\sim 1/\chi_0 \sim \xi^{-2}$, we find, in the limit $k\ra \xi^{-1}$, $\omega\ra \xi^{-z}$:
\beq \begin{split}
\Pi_{u} &\propto \xi^{8-2d},\quad  \Pi_{u'} \propto \xi^{10-3d}\,,\quad \Pi_{r,\kappa,\nu_l,j} \propto \xi^{2-d}\,.
\end{split}\label{scal_mf}\eeq
The identical scaling of all one-loop diagrams of Fig.~\ref{fig:self_en_dyn} is a consequence of the identical scaling behavior of the three-point vertices in the model.
We can also obtain scaling predictions beyond the mean-field case by making use of our knowledge of the proper critical behavior of the static couplings, $r$, $\kappa$ and $u$. These are renormalized by fluctuations as $r_R\sim \xi^{-2+\eta}$, $\kappa_R\sim \xi^{\eta}$, $u_R\sim \xi^{d-4+2\eta}$ \cite{ma_book, onuki_book}, implying that
\beq \Pi_u\propto \xi^{z-2+\eta}\,,\quad \Pi_{u',r,\kappa,\nu_l}\propto \xi^{z-d}\,.
\label{scal_crit}
\eeq
The contributions due to $\Pi_j$ all scale $\propto \xi^{z-d\pm O(\eta)}$, or $\xi^{z_t-d\pm O(\eta)}$, respectively, up to differences in the exponent of $O(\eta)$ accounting for possible divergences of $\nu_l$ or $\nu_t$. Thus, all diagrams except $\Pi_u$ are irrelevant for $d>2$, provided that $z$ and $z_t$ are still close to 2, which will indeed be the case. 
Note that the contributions from the $u\phi^4$-vertex, which were found to diverge below a critical dimension of $d=10/3$ in the mean-field limit, now remain finite at least down to three dimensions, due to the renormalization of $u$.
We also see that the $\jv/\rho$-nonlinearity in eq.~\eqref{nse1}, responsible for the coupling between longitudinal and transverse current, is not relevant for $d>2$ and it is safe to approximate $\rho$ by $\rho_0$, as far as asymptotic critical properties are concerned. 

From the dominance of $\Sigma_u$, which arises from the $\phi^3$-vertex of $p\st{nl}$, we conclude that the \textit{upper critical dimension} of the present isothermal non-ideal fluid model is $d_c=4$ \textit{both} in statics and dynamics.
The present analysis also reveals that the relevant nonlinearities responsible for the deviations from the classical (van Hove) predictions are different for the isothermal fluid and model H: in the latter case, the deviation is caused by the reversible advection term [see eq.~\eqref{modelH_cont}], whereas in the isothermal fluid, it is caused by the dissipative $\phi^4$-nonlinearity of the Ginzburg-Landau free energy functional. Hence, in the isothermal fluid, the dominant dynamic critical effects are induced by quantities of purely thermodynamic origin.

The scaling result for $\Pi_u$ of eq.~\eqref{scal_crit} is not sufficient to determine the precise value of $z$. This can be done via a renormalization group (RG) calculation slightly below four dimensions. To this end, the wavenumber integrations in $\Pi_u$ of eq.~\eqref{self_u2} are performed incrementally in a shell $\Lambda_0 e^{-s}<q,q'<\Lambda_0$, where $\Lambda_0$ is a cutoff and $s$ denotes the RG flow parameter. The contribution linear in $s$, which we shall write as $A(\nu_l)\,\nu_l$, has been calculated in \cite{folk_schwabl_elastic_prb1979}, with the essential result that $A(\nu_l\ra \infty)=6 \ln(4/3)\eta$ and $A(\nu_l)>A(\infty)$ for any finite $\nu_l$. These estimates have been obtained at $O(\epsilon^2)$ in an $\epsilon$-expansion.
The RG equation for the longitudinal viscosity then reads
\beq \pd_s \nu_l(s) = A(\nu_l) \nu_l(s)\,,
\eeq
from which one concludes that, for any positive bare $\nu_l(0)$, $\nu_l(s)$ will grow along the RG flow and asymptotically scale as $e^{A(\infty)s}$. Thus, in the hydrodynamic limit, which is reached for $e^s\sim \Lambda_0\xi$ \cite{onuki_book, kopietz_frg_book}, the renormalized longitudinal viscosity behaves as
\beq \nu_{lR} \propto \nu_l \xi^{x}\,,
\label{long_visc_rg}
\eeq
with the critical index being
\beq x=6\ln(4/3)\eta\simeq 1.7\eta\,.
\label{visc_index}
\eeq
In the critical regime ($k\gg \xi^{-1}$), we have accordingly, $\nu_{lR}\propto \nu_l k^{-x}$.

With this result we can show that sound waves must be \textit{overdamped} in the critical isothermal fluid: using the fact that $c_{sR}^2\propto \xi^{-2+\eta}$, the linear hydrodynamical condition for strong damping, 
$c_{sR}(k)\ll k\nu_{lR}\,,$
becomes
$\xi^{-1+\eta/2}/k + \text{const.}\times k^{-\eta/2} \ll \xi^{x}$. Thus, for wavenumbers of order $k\sim \xi^{-1}$, we have
$\xi^\eta \ll \xi^{2x}$, which is always fulfilled in the asymptotic critical regime since $2x>\eta$.
Of course, we could also have kept, for $d$ close to $d_c$, only the dominant $u\phi^3$-nonlinearity in eq.~\eqref{crit_den_eq} and thereby recover the type of sound-wave equation studied in the context of isotropic elastic phase transitions \cite{folk_schwabl_elastic_prb1979}. The associated RG analysis in \cite{folk_schwabl_elastic_prb1979} then leads to the same predictions as above. 

Concluding, in the critical regime, eq.~\eqref{crit_den_eq} reduces in the long-time limit to \textit{model A} in the classification of \cite{halperin_hohenberg_ma_prl1972, halperin_hohenberg_ma_prb1974, halperin_hohenberg_ma_prb1976, hohenberg_halperin}, that is, a time-dependent Ginzburg-Landau model for a non-conserved order parameter of the form
\beq - \im \omega \drho = \frac{1}{\nu_l} \frac{\delta \Fcal}{\delta\rho} + \mathcal{R} + h_l\,,  \label{crit_den_od}\eeq
where $\mathcal{R}\equiv R_l/\nu_l$ is a noise source of variance $\sim 1/\nu_l$ \footnote{The conserved nature of the fluid order parameter (density) becomes noticeable at early times, where the correlation function decays non-exponentially and the dynamics deviates from pure model-A behavior.}.
Since overdamped sound waves relax with a rate
\beq \Gamma  = \frac{c_{sR}^2 }{\nu_{lR}} \,,
\label{dyn_crit_damp}
\eeq
the dynamic critical index $z$, defined via the relation $\Gamma \propto \xi^{-z}$, of the fully nonlinear fluid model follows as 
\beq z=2-\eta+x\,.
\label{z_dyn}
\eeq 
In contrast, in the linear hydrodynamic (mean-field) case, we have $z=2$ and $x=0$. 
If we assume that the pure model-A behavior of the critical isothermal fluid persists also in low dimensions, we expect, in the interesting two-dimensional case, a value of 
\beq z\simeq 2.08 \ldots 2.17\quad \text{(2D)}\,,
\label{z_est}\eeq
based on recent theoretical calculations and Monte Carlo simulations of model A \cite{krinitsyn_prudnikov_ising_2006, canet_nprg_modelA_jphysA2007,nightingale_bloete_PRL1996}. 
In the wider literature, varying estimates for $z$, ranging between $2.0$ and $2.3$, have been reported \cite{rogiers_indekeu_prb1990,dunlavy_venus_prl2005}.
Above value for $z$ translates to $x\simeq 0.4$ and agrees surprisingly well with the $O(\epsilon^2)$-renormalization-group prediction of eq.~\eqref{visc_index} in 2D.
For comparison, for a conventional fluid (model H), we have $z\simeq d$ and $x=z-\alpha/\nu$ \cite{hohenberg_halperin, onuki_pre1997, onuki_book}.

Returning to the scaling estimates of eq.~\eqref{scal_crit}, a value of $z>2$ would imply the weak divergence of various diagrams in 2D, which could provide corrections to the critical exponents. 
To address this issue, explicit calculations of the corresponding contributions will be required. Our simulations (see sec.~\ref{sec:sim}) yield a value of $z\approx 2.2\pm 0.1$, suggesting that possible corrections to the model-A behavior are small at least.

\subsection{Critical shear viscosity}
\label{sec:theo_viscs}
The shear viscosity is computed in the following based on a Green-Kubo approach. 
We consider the $x$-component of the nonlinear NSE \eqref{nse1} and choose the wavevector to lie along the $y$-direction, i.e.\ $\kv=(0,k)$ in 2D.
Applying the approximation $\jv/\rho \simeq \jv/\rho_0 - (\jv/\rho_0^2) \drho$ to the viscous stress tensor, whose $xy$-component becomes
\beq S_{xy} = \nu_s[ \pd_x(j_y - j_y \phi) + \pd_y(j_x-j_x \phi)]
\label{visc_stress_ten}
\eeq
the equation for the transverse current can be written as
\beq \pd_t j_x = - \nu_s k^2 j_x -\im k P_{xy}(\kv) - \im k S\ut{nlin}_{xy}(\kv) + \im k R_{xy}\,,
\label{jx_nonlin_nse}
\eeq
where $P_{xy}(\kv,t)= - \kappa \int_\qv q_x q_y\, \phi(\qv) \phi(\kv-\qv) $ is the Fourier-transform of the off-diagonal term of the thermodynamic pressure tensor [eq.~\eqref{press_ten}], while $S\ut{nlin}_{xy}(\kv,t) = \im \nu_s k \int_\qv j_x(\qv,t) \phi(\kv-\qv,t)$ contains the nonlinear terms of $S_{xy}$ [eq.~\eqref{visc_stress_ten}] involving the order parameter.

The fluctuation contribution to the shear viscosity, $\nu_{s,\text{crit}}$, can now be inferred by invoking a Green-Kubo relation \cite{perl_ferrel_pra1972, hao_viscosity_pre2005, onuki_book}. For the contribution from the thermodynamic pressure tensor we find 
\beq\begin{split}
\nu_{s,\text{crit}} &= \frac{1}{V  k_B T} \lim_{k\ra 0} \int_0^\infty dt\, \bra P_{xy}(\kv,t) P_{xy}(-\kv,0)\ket \simeq \frac{\kappa^2}{k_B T} \int_\qv q_x^2 q_y^2 \frac{C^2(\qv)}{\Gamma(\qv)} \propto \xi^{z-d} \,,
\label{shear_visc1}
\end{split}\eeq
where, as usual, the four-point correlation has been decoupled into products of two-point correlation functions. 
For the contribution due to the nonlinear part of the viscous stress tensor, $S_{xy}$, one writes $j_x(\qv)=j_l(\qv)\hat q_x + \jv_{t,x}(\qv)$ and uses the fact that the correlation function of the longitudinal current fulfills $C_l(\qv,t)=\pd_t^2 C(\qv,t)/q^2$. Also, $j_l$ and $\jv_t$ are independent to leading order. This gives analogously
\beq\begin{split}
\nu'_{s,\text{crit}} &= \frac{1}{V  k_B T} \lim_{k\ra \xi^{-1}} \int_0^\infty dt\, \bra S\ut{nlin}_{xy}(\kv,t) S\ut{nlin}_{xy}(-\kv,0)\ket \propto \xi^{4-z-d-2\eta} + \xi^{z_t-\eta-d}\,,
\label{shear_visc2}
\end{split}\eeq
where the external wavevector is taken at $\xi^{-1}$ and, for the evaluation of the part involving the transverse current, it has been assumed that $z_t< z$.
In the mean-field limit, all contributions scale $\propto \xi^{2-d}$ and thus are finite for $d>2$, while a potential logarithmic divergence is indicated in 2D. If, in contrast, scaling exponents appropriate for the true critical point are taken (where $z\simeq 2.2$ and $\kappa_R\propto \xi^\eta$), the contribution to $\nu_{s,\text{crit}}$ of eq.~\eqref{shear_visc1} attains a weak power-law divergence in 2D, characterized by a critical exponent $y=z-2$, implying that $z_t=2-y<2$ in 2D. As a consequence, eq.~\eqref{shear_visc2} becomes now finite in all dimensions.
Note also that the order-parameter self-energy is very sensitive to a possible divergence of the shear viscosity, as the associated scaling analysis suggests [cf.~eq.~\eqref{scal_crit}]. Clearly, at this stage more detailed calculations are needed to obtain quantitative predictions for the dynamic critical exponents in 2D.
The same calculation as in eq.~\eqref{shear_visc1} applies also to model H, consistent with the well-known RG result $y=z-d$ \cite{siggia_prb1976, hohenberg_halperin, kawasaki_review1976, onuki_book} (see Tab.~\ref{tab:crit_fluid}). In model H, the shear viscosity diverges mildly in 3D, due to a value of $z\simeq 3.07$ that is slightly larger than 3 \cite{hohenberg_halperin, hao_viscosity_pre2005}.  In 2D, the uncertainty in the theoretical value for $z$ (see Tab.~\ref{tab:crit_fluid}) and the lack of experimental or numerical studies permits no definite conclusion on a possible critical divergence of the shear viscosity in an ordinary fluid.


\section{Simulations}
\label{sec:sim}
The theoretical predictions are now compared to full fluctuating hydrodynamics simulations of an isothermal non-ideal fluid in 2D, using the Lattice Boltzmann (LB) model introduced in \cite{gross_flbe_2010}. 
For a brief description of the simulation model, specifically in regard to critical phenomena, we refer to \cite{gross_critstat_2012}.
There, also the static critical behavior of the system, which belongs to the 2D Ising universality class, is analyzed. 
The two-dimensional case is interesting for several reasons: first, the isothermal condition could probably be realized here most easily experimentally by coupling the fluid to a heat-absorbing substrate. Second, the scaling arguments of sec.~\ref{sec:theor} suggest that the fluctuation contributions of various nonlinearities grow around a dimension of $d=2$. As such effects are complicated to assess analytically, numerical simulations can provide useful insights and are complementary in this case.

\subsection{Setup}
Parameters of our LB simulations are chosen as in \cite{gross_critstat_2012}, a typical setup at the critical point being $r= -4.8\times 10^{-5}$, $u=2.8 \times 10^{-2}$ and $\kappa=9.6\times 10^{-5}$. This choice leads to a mean-field interface width of $\simeq 2$ lattice units (l.u.) and is expected to ensure reliable results on the fluctuation dynamics \cite{wagner_interface_2007, gross_critstat_2012}. The average density in our simulations is $\rho_0=1.0$. The noise temperature is set to $k_B T=10^{-7}$ and the bare shear and bulk viscosities to $\nu_s=\nu_b/2=0.04/3$.
Quantitatively similar results have been obtained also for other parameter combinations.
Simulation boxes are of size $L\times L=256^2$, except for Figs.~\ref{fig:relax} and \ref{fig:viscs}c, where $L\times L=128^2$.
All simulation results reported in the paper are obtained with a standard LB implementation, where the viscous stress consists of terms of the form $\nu\rho\pd_\alpha u_\beta$, i.e., the dynamic viscosities depend on $\rho$.
In a few cases it has been checked, by using an implementation where the $\rho$-dependence of the dynamic viscosities is eliminated, that results are not affected by this LB specific peculiarity. Besides, due to requirements of numerical stability, the relative density fluctuations $\delta\rho/\rho_0$ in our simulations are on average well below a few percent, thus warranting the approximation $\rho\simeq \rho_0$.
Due to the multiplicative nature of the renormalization of the relaxation rate, the specific values of the viscosities are not important in this regard. They do, however, influence the extension of the overdamped acoustic regime and the crossover from mean-field to the expected model-A critical behavior.
While a small longitudinal viscosity leads to a more rapid equilibration of the order parameter, it also shifts the onset of the overdamped regime to smaller wavenumbers.
As a consequence, larger simulation boxes would be required to reduce the residual speed of sound at the critical point sufficiently.
Furthermore, a large value of $c_s$ can significantly affect the long-time decay of the order-parameter correlation function, which can be misinterpreted as caused by a larger viscosity.
In order to diminish these and other undesired finite-size effects, the lowest $k$-modes are usually excluded from the analysis of our results.
To avoid effects of lattice anisotropy (cf.\ \cite{sumesh_derivative_2012}), all wavenumber-dependent quantities shown in the plots are computed as an average over the Cartesian axes of the Fourier plane.

The logarithmic divergence of the viscosities in 2D due to the convective nonlinearity \cite{pomeau_pra1972, pomeau_mct_review1975, forster_nelson_prl1976, forster_nelson_pra1977} is difficult to observe and requires either very long simulation times or large simulation boxes (cf.~\cite{frisch_complexsys1987}). 
Indeed, since the effect is proportional to $k_B T \log L$ \cite{mazenko_book}, it is expected to be below the threshold of statistical accuracy for the present setup. 

\subsection{Results}
\subsubsection{Order parameter}

\begin{figure*}[t]
(a)\includegraphics[width=0.45\linewidth]{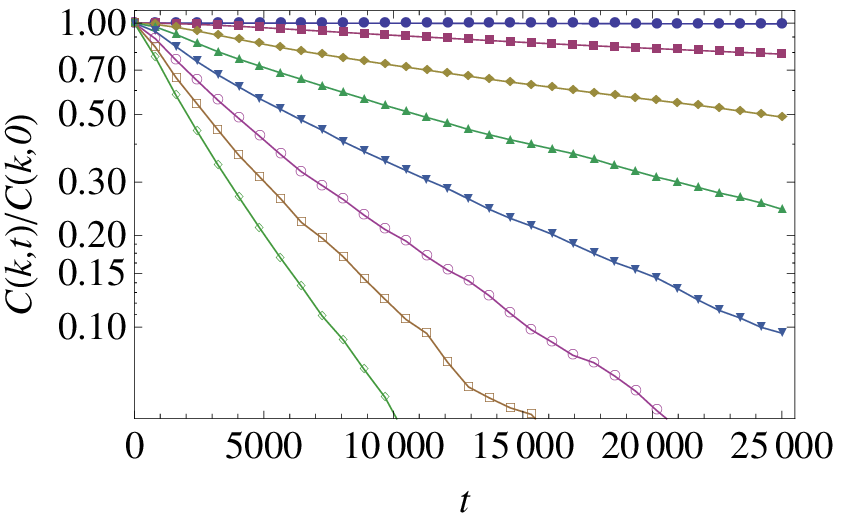}
(b)\includegraphics[width=0.43\linewidth]{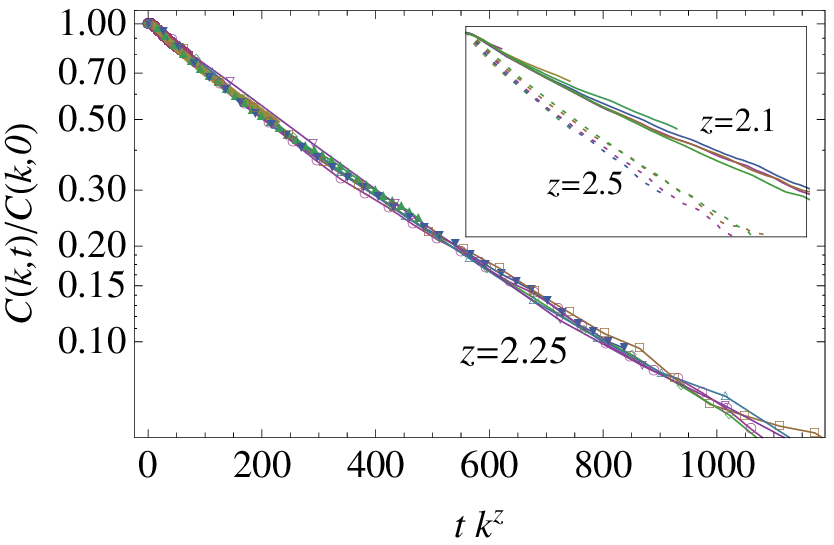}
(c)\includegraphics[width=0.45\linewidth]{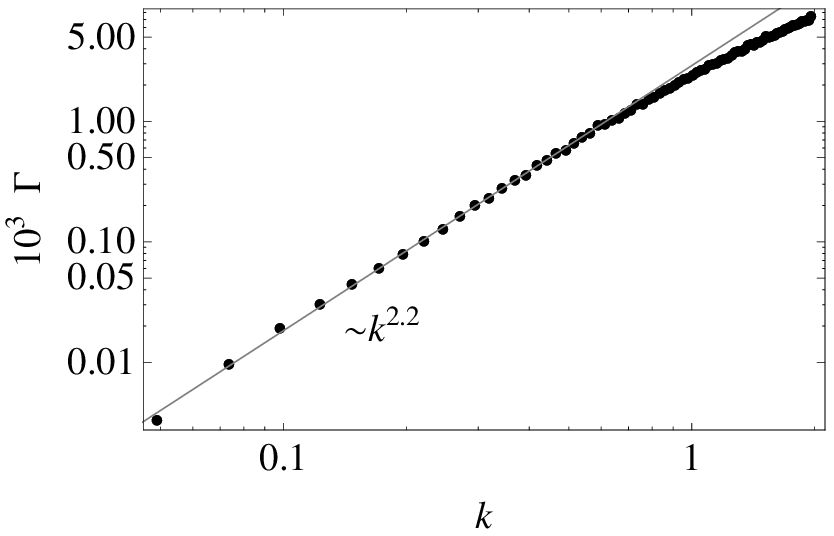}
\caption{(a,b) Dynamic structure factor at the critical point: (a) Raw data for the lowest wavenumbers $k$ (increasing from top to bottom). The time is given in lattice units and the lines are drawn as a guide to the eye. 
(b) Test of the dynamic scaling form of $C(k,t)$. The value of the dynamic critical index $z$ is determined by varying $z$ until all points fall onto a master curve. The inset shows that the collapse is incomplete for $z$ significantly different from $2.25$. (c) Relaxation rate $\Gamma$ at the critical point obtained from exponential fits to the dynamic structure factor for different wavenumbers. The solid line is $\propto k^{2.2}$. 
}
\label{fig:sfdyn}
\end{figure*}

Fig.~\ref{fig:sfdyn}a shows the dynamic structure factor $C(k,t)$ at the critical point for different wavenumbers $k$ as obtained from our simulations.
The exponential decay of $C(k,t)$ is clearly seen in the semi-logarithmic representation.
According to the dynamic scaling hypothesis,
$C(k,t) = k^{-2+\eta} \mathcal{C}((k \xi)^{-1}, k^z t)$;
hence, sufficiently close to the critical point, where $(k \xi)^{-1}$ is small, the dynamic critical index $z$ can be determined by plotting $C(k,t)/C(k,0)$ versus the rescaled time $k^z t$, testing different values of $z$ until a complete data collapse is achieved.
This is done in Fig.~\ref{fig:sfdyn}b, from which we infer a value of $z\simeq 2.25\pm 0.1$. For comparison, the insets demonstrate that, when rescaling the data with a significantly larger or smaller value of $z$, the data collapse remains incomplete.

Alternatively to the rescaling procedure, the dynamic critical index can be more directly determined from the relaxation rate $\Gamma(k)$, which can be obtained by fitting an exponential decay [eq.~\eqref{dyn_correl_time}] to the dynamic structure factor.
The assumption of an exponential relaxation is well satisfied in the overdamped regime, after neglecting the short-time, non-exponential part of $C(k,t)$ caused by a finite residual speed of sound.
In Fig.~\ref{fig:sfdyn}c, the so obtained relaxation rate is plotted against the wavenumber in a double-logarithmic representation. At small wavenumbers, the expected power-law behavior $\Gamma\propto k^z$ is clearly seen. A numerical fit yields a value of the exponent of $z\simeq 2.2\pm 0.2$, which agrees well with the value obtained from rescaling the structure factor data.

These results show that the dynamic critical index is significantly increased over its mean-field ($z=2$) or van Hove ($z=1.75$) value. 
In particular, the extracted value of $z$ is consistent with the presence of pure model-A-type critical behavior [eq.~\eqref{z_est}] in two dimensions.

We remark that the value of $z$ depends in principle also on the range of wavenumbers considered.
The deviation from a pure power-law at larger wavenumbers might be caused by the general wavenumber-dependence of the LB transport coefficients \cite{behrend_hydro_1994, lallemand_theory_2000} and the influence of other nonlinearities in the model that overwhelm the leading order critical divergences. For $k\gtrsim 1$ also the discrete nature of the lattice becomes noticeable (see, e.g., \cite{gross_flbe_2010, gross_critstat_2012}).
More precise values for $z$ could be obtained by using larger simulation boxes, thereby extending the low-$k$ regime and decreasing the influence of the non-exponential decay of $C(k,t)$ at short times.

\begin{figure*}[t]
(a)\includegraphics[width=0.46\linewidth]{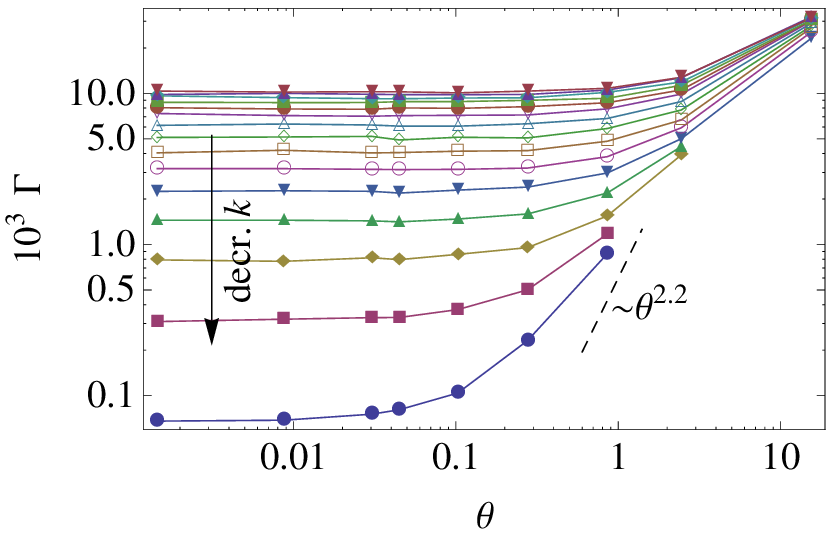}
(b)\includegraphics[width=0.46\linewidth]{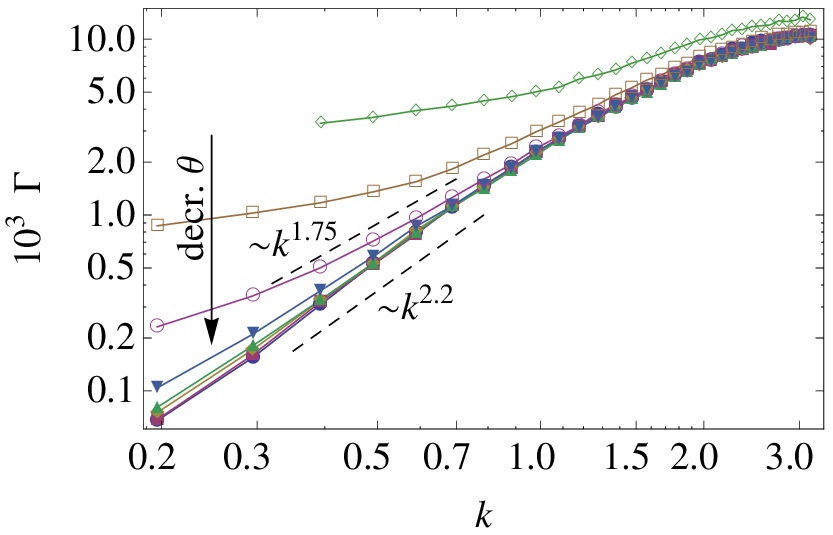}
(c)\includegraphics[width=0.45\linewidth]{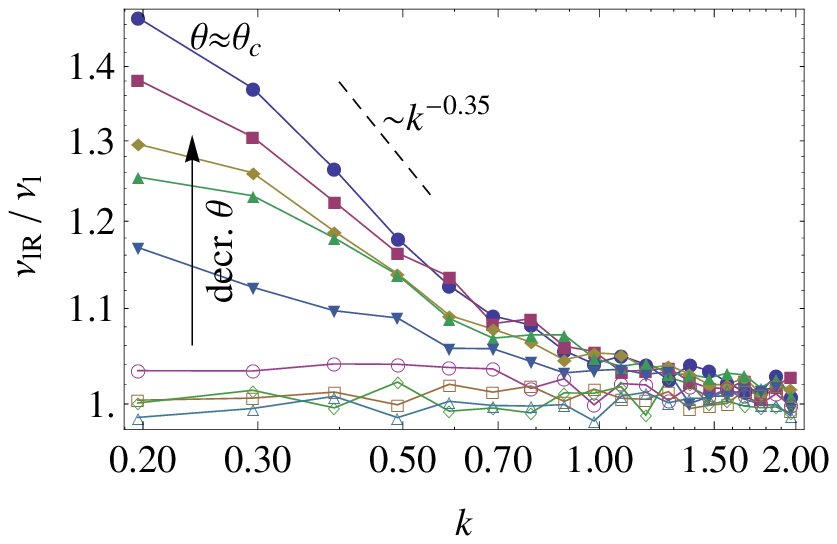}
\caption{(a) Dependence of the relaxation rate on the reduced temperature $\theta=(r_c-r)/r_c$ ($r$ is the Landau parameter and $r_c$ its value at the critical point) for different wavenumbers. The expected power-law decrease of $\Gamma$ (dashed line) is rounded off when approaching the critical temperature due to the finite system size. (b) Wavenumber-dependence of the relaxation rate for different reduced temperatures approaching the critical point from above. Data points in the transition region between the overdamped and propagating acoustic regime are omitted. The power-law $\sim k^{2-\eta}\sim k^{1.75}$ corresponds to the van Hove prediction for $\Gamma$, where a wavenumber-independent $\nu_{lR}$ is assumed. (c) Effective longitudinal viscosity $\nu_{lR}$ (normalized to its bare value $\nu_l$) in dependence of temperature and wavenumber, approaching the critical point from above. Finite-size effects, leading to a weaker-than-expected divergence of $\nu_{lR}$, are noticeable at low $k$ (see text). The lines are drawn as a guide to the eye. 
}
\label{fig:relax}
\end{figure*}

In Fig.~\ref{fig:relax}, the behavior of the order-parameter relaxation rate (obtained from exponential fits to the dynamic structure factor) is investigated in greater detail for different wavenumbers $k$ and reduced temperatures $\theta$, where $\theta=(r_c-r)/r_c$.
As seen in Fig.~\ref{fig:relax}a, the relaxation rate at fixed $k$ first markedly decreases upon approaching the critical point ($\theta=0$), but eventually levels off at a finite value due to a nonzero speed of sound.
Also, the expected temperature dependence $\Gamma\propto \theta^{z\nu}$ with $z\nu\simeq 2.2$ is not observed, but instead a less steep decrease. These effects are well known consequences of the finite system size and are observed also for static quantities (see \cite{gross_critstat_2012}). 
Analogously to statics, finite-size effects can be expected to be much less pronounced when looking directly at the wavenumber dependence of a critical quantity.
Indeed, in Fig.~\ref{fig:relax}b it is clearly seen that, when approaching the critical point, $\Gamma$ smoothly assumes its expected power-law $\propto k^z$.
Sufficiently far above the critical point, order-parameter modes at low $k$ cross over to the propagating regime and a relaxation rate can no longer be defined.

In Fig.~\ref{fig:relax}c, the effective longitudinal viscosity, $\nu_{lR}(k)=c_{sR}^2(k)/\Gamma(k) =  k_B T/[\Gamma(k) C(k)]$, computed from the data of the relaxation rate and static structure factor, is shown.
Far above the critical temperature, $\nu_{lR}$ assumes its bare value $\nu_l$ and is practically independent of wavenumber, while at criticality ($\theta=0$), the expected power-law divergence $\nu_{lR}\propto  k^{-x}$ is reproduced with reasonable accuracy. Note that the critical enhancement is multiplicative and independent of the bare viscosity.
Deviations at the lowest $k$ can be attributed to the relatively strong finite-size effects that occur in the static structure factor of the present model \cite{gross_critstat_2012}: at low $k$, $C(k)$ appears slightly steeper than the expected $k^{-2+\eta}$-power law at the critical point, which is reflected in a weaker-than-expected divergence of the longitudinal viscosity.

\subsubsection{Shear viscosity}
\label{sec:sim_shear}

\begin{figure}[t]
\centering
(a)\includegraphics[width=0.41\linewidth]{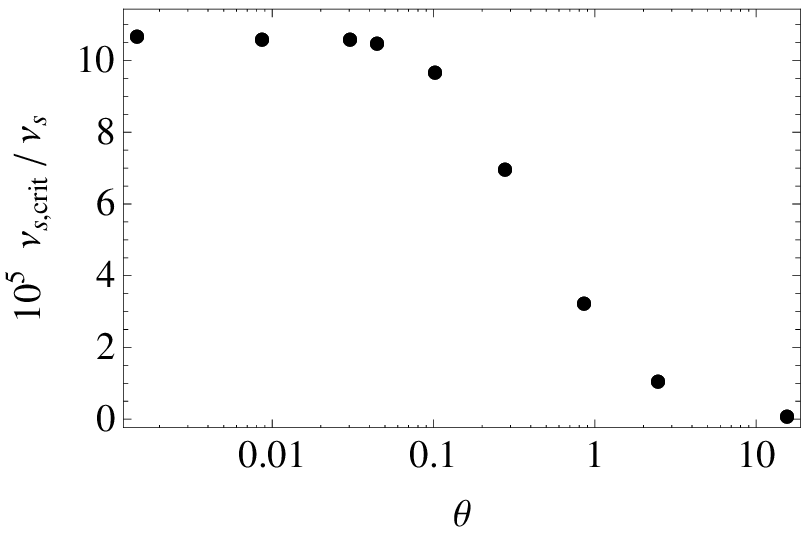}
(b)\includegraphics[width=0.43\linewidth]{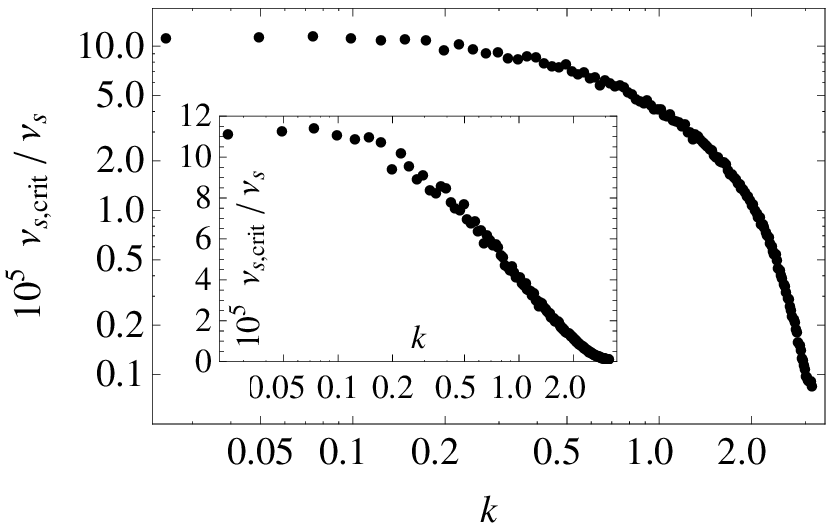}
(c)\includegraphics[width=0.44\linewidth]{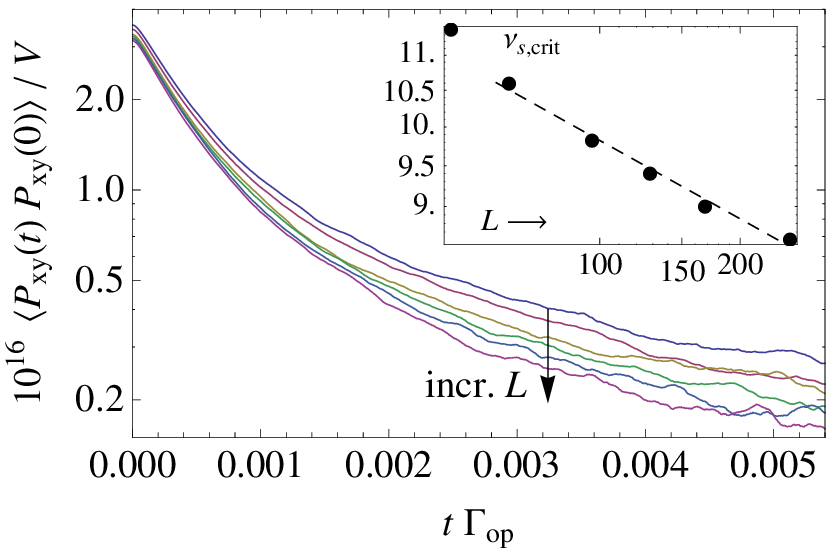}
\caption{(a) Critical fluctuation contribution to the shear viscosity in dependence of the reduced temperature. The data (filled symbols) is obtained from simulations using the Green-Kubo relation \eqref{shear_visc_sim} at $k=0$. The solid curve represents the theoretical prediction of eq.~\eqref{shear_visc1}. (b) Wavenumber-dependence of the critical shear viscosity (at $\theta=0$) obtained from eq.~\eqref{shear_visc_sim}, plotted in double-logarithmic representation. For comparison, the same data is plotted in the inset in log-linear scale. (c) Finite-size behavior of the shear stress correlation function (normalized by the system volume $V=L^2$) and the critical fluctuation contribution to the shear viscosity, $\nu_{s,\text{crit}}$ (inset). The dashed curve is $\propto L^{-0.15}$. Time is expressed in units of the characteristic order-parameter relaxation time $1/\Gamma\st{op}$. The theoretically predicted power-law divergence of the shear viscosity is not recovered by the present simulations. See text for further discussion.}
\label{fig:viscs}
\end{figure}

Turning to the critical behavior of the shear viscosity, we study here only the contribution from the thermodynamic pressure tensor, $P_{xy} = \kappa(\pd_x \rho) (\pd_y \rho)$, via the Green-Kubo relation 
\beq \nu_{s,\text{crit}}(\kv) =\int_0^\infty dt\, \bra P_{xy}(\kv,t) P_{xy}(-\kv,0)\ket/ (V  k_B T)
\label{shear_visc_sim}
\eeq [cf.~eq.~\eqref{shear_visc1}]\footnote{See also \cite{ladd_1994, laddVerberg_2001} for an application of the Green-Kubo formalism to determine viscosities in the Lattice Boltzmann method.}.
Fig.~\ref{fig:viscs}a shows the so obtained temperature-dependence of $\nu_{s,\text{crit}}$ at $k=0$ close to the critical point, whereas Fig.~\ref{fig:viscs}b shows the wavenumber-dependence of $\nu_{s,\text{crit}}$ at the critical temperature ($\theta=0$) on a double-logarithmic (main plot) and logarithmic-linear (inset) scale. 
While there appears some logarithmic growth of $\nu_{s,\text{crit}}(k)$ at larger $k$, the plateau at low $k$ suggests that the shear viscosity stays finite in 2D, in disagreement with the scaling predictions of eq.~\eqref{shear_visc1}. Note that, although the shear-viscosity at $k=0$ when plotted against reduced temperature (Fig.~\ref{fig:viscs}a) exhibits an extended plateau towards $\theta\ra 0$ as well, this effect can not be unambiguously attributed to the non-divergent nature of $\nu_{s,\text{crit}}$, as finite-size effects are expected to contribute significantly here (cf.~Fig.~\ref{fig:relax}a, where a similar effect is seen for the relaxation rate).
The correlation function of the thermodynamic shear stress, $\bra P_{xy}(t), P_{xy}(0)\ket$, is found to decay in a non-exponential manner over a characteristic timescale that is much shorter than the relaxation of the order parameter (Fig.~\ref{fig:viscs}c).
The finiteness of the critical shear viscosity can also be inferred from its scaling behavior with the system size $L$ (inset to Fig.~\ref{fig:viscs}c). Interestingly, $\nu_{s,\text{crit}}$ is found to even decrease with larger $L$ by a power-law with a small exponent of roughly $-0.15$. From the system-size dependence of the shear-stress correlation function, it is concluded that this behavior arises from both a decrease of the shear-stress relaxation time and a decrease of the equal-time autocorrelation of $P_{xy}$ with $L$ (main plot of Fig.~\ref{fig:viscs}c). We find a similar behavior also slightly away from the critical point, although the effect is less pronounced there. 

A possible reason for the disagreement with the critical scaling predictions of eq.~\eqref{shear_visc1} might be that, by computing the correlation function of the stress tensor $P_{xy}$ in our simulations, the renormalization of the square-gradient parameter $\kappa$ is not properly taken into account, since it enters here only as a constant numerical prefactor. Indeed, if this effect is neglected in eq.~\eqref{shear_visc1}, the scaling exponent changes to $z-d-2\eta<0$ in 2D, implying a non-divergent $\nu_{s,\text{crit}}$. An alternative method to determine the effective shear viscosity would be to compute the transverse current correlation function. However, due to the small value of $\kappa$, which is a necessity of the present LB model (see \cite{gross_flbe_2010, gross_critstat_2012}), $\nu_{s,\text{crit}}$ remains orders of magnitude below its bare value $\nu_{s}$. Thus, exceeding computational resources would be necessary to extract the fluctuation contribution to the shear viscosity from the transverse current correlation function. Clearly, a different critical behavior of the shear viscosity can also have repercussions on the order-parameter dynamics and lead to a slightly different value of the dynamic critical exponent $z$. Further numerical investigation of the shear viscosity in a two-dimensional isothermal fluid using alternative simulation methods are thus desirable for future work.

\section{Discussion}
In a conventional single-component fluid (model H), the dominant transport mechanism is heat diffusion, while sound waves are decoupled from the order-parameter dynamics \cite{stanley_book, kawasaki_review1976, onuki_book, bhattacharjee_sound_rpp2010}. 
In contrast, under isothermal conditions, heat diffusion is absent and order-parameter fluctuations can relax only via sound waves. 
Based on scaling considerations it has been argued here that, below four dimensions, the order-parameter dynamics of the critical isothermal fluid is characterized by model-A-type behavior.
This implies that, at long wavelengths, sound waves are overdamped due to a strongly diverging bulk viscosity, $\nu_l\propto \xi^x$. 
The relaxation rate, $\Gamma=c_s^2/\nu_l$, scales as $\xi^{-z}$ with a dynamic critical exponent of $z=\gamma/\nu+x=2-\eta+x$, where $\gamma/\nu$ represents the contribution from the isothermal speed of sound and $x\approx 1.7\eta$ represents the ``non-classical'' contribution from the renormalized bulk viscosity. 
While in model H, the divergence of the kinetic coefficient arises due to the advective coupling to the transverse velocity modes \cite{kawasaki_review1976, onuki_book}, it is due to the nonlinear thermodynamic pressure in the isothermal case.
Longitudinal and transverse currents are approximately decoupled for $d$ close to 4 -- a property that becomes exact in the linear case.
In 2D, our simulations of the fluctuating hydrodynamic equations yield a value of $z\approx 2.2\pm 0.1$ and $x\approx 0.45\pm 0.1$, in reasonable agreement with theoretical expectations and Monte-Carlo simulations of model A.
The scaling theory predicts the shear viscosity to remain finite for $d>2$ and weakly diverge by a power-law in two dimensions. This divergence could, however, not be observed within the present simulation approach.
The essential differences in the critical dynamics of an isothermal and an ordinary fluid (model H) are collected in Tab.~\ref{tab:crit_fluid}.

It is interesting to compare the present findings also to the situation in hydrodynamic models of the glass transition \cite{das_mazenko_prl1985, das_mazenko_pra1986, kawasaki_mct_review_1995, das_book}, where the isothermal compressible Navier-Stokes equations have been investigated in conjunction with a purely Gaussian free energy. There, the density correlation function shows an an anomalously slow decay at low temperatures, accompanied by a strong increase of the bulk viscosity. Quite analogously to the case in critical dynamics, this is a generic mode-coupling effect caused by a nonlinear pressure term. In the case of a supercooled liquid, however, the dominant contribution arises from the quadratic pressure nonlinearity [$r\phi^2$ in eq.~\eqref{pnl}], whereas in the critical fluid, this term turns out to be irrelevant due to the smallness of $r$.

The present model is particularly interesting in the two-dimensional case, where an experimental realization of the isothermal condition might be achievable. While the present scaling considerations indicate the divergence of additional contributions to the bulk viscosity beyond the model-A term, our simulation results suggest that the model-A-type critical behavior essentially persists also in 2D, with possible corrections to critical exponents being small, at least.
For future work, it will thus be interesting to treat the isothermal fluid model within a renormalization-group approach and derive more detailed predictions in the two dimensional case. 
In order to clarify the critical behavior of the shear viscosity in 2D, alternative simulation methods could be invoked.

The predictions obtained in this work might be experimentally testable on single-component monolayer films that admit for liquid-vapor-like phase-separation below a critical point \cite{knobler_sci1990, kaganer_monolayer_review1999}.
Of course, the present model is highly idealized in that it neglects the possible influence of electrostatic long-range interactions \cite{mcconnell_monolayers_review1991, folk_moser_longrange_pre1994, binder_luijten_longrange_physrep2001, belim_longrange_jetp2004}, friction between fluid and substrate, and hydrodynamic back-coupling \cite{ramaswamy_mazenko_pra1982, lubensky_goldstein_physfl1996, alexander_langmuir_jfm2007, haataja_pre2009}.
Also, it is assumed that the rate of heat transfer between fluid and substrate is sufficiently large to provide an effective isothermal environment for the critical fluctuations (cf.~\cite{griesbauer_wixforth_bioj2009}). 
Since the long-wavelength dynamics of a fluid becomes arbitrarily slow upon approaching the critical point, one might expect that even a relatively small thermal coupling will actually be sufficient. 

From a theoretical perspective, crossover behavior between different dynamic universality classes of a single-component fluid film is expected: in case of negligible friction, model H is obtained for vanishing thermal coupling and, as shown in this work, model A for perfect thermal coupling. In the opposite case of large friction, it is expected that model B (i.e., a purely diffusive order-parameter transport) results \cite{ramaswamy_mazenko_pra1982, casalnuovo_pra1984}.

\begin{acknowledgements}
We would like to thank R. Adhikari, M. E. Cates, H. W. Diehl, S. May and A. J. Wagner for helpful discussions. Funding from the DFG (Va205/5-3), the industrial sponsors of ICAMS, the state of North-Rhine Westphalia and the European Commission in the framework of the European Regional Development Fund (ERDF) is gratefully acknowledged. M.G. would also like to thank the EPCC, Edinburgh and the NDSU, Fargo for hospitality.
\end{acknowledgements}

\appendix
\section{Further contributions to the self-energy}
\label{app:add_diagr}

\begin{figure}[t]
\centering
\includegraphics[width=0.96\linewidth]{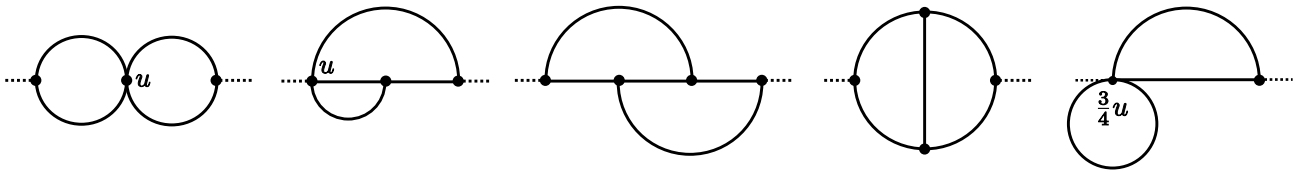}
\caption{Schematic diagrams of the additional contributions to the order-parameter self-energy [cf.~eq.~\eqref{gfull_selfen}] at two-loop order. Unlabeled dots represent the possible couplings $r$, $\kappa$, $\nu_l$, $\nu_t$ associated with the three-point vertices. Open circles and arrows (indicating correlation or response functions) are understood to be present on some of the internal lines. }
\label{fig:add_diagr}
\end{figure}
The contributions to the order-parameter self-energy at two-loop order are diagrammatically shown in Fig.~\ref{fig:add_diagr}.
We have omitted diagrams where some of the solid lines are replaced by wavy lines representing transverse current response/correlation functions.
Regarding the scaling behavior, these types of diagrams need not be explicitly evaluated, since all three-point vertices scale in the same way.
Taking into account the proper critical behavior of the couplings, the dynamic part of each diagram in Fig.~\ref{fig:add_diagr} is found to scale $\propto \xi^{2-d}$ (up to corrections of exponents of $O(\eta)$) and thus gives negligible contributions to the renormalized viscosity for $d>2$.


\end{document}